\documentclass[12pt]{iopart}
\usepackage{graphicx}
\usepackage{color}

\def\bmu{\mbox{\boldmath{$\mu$}}}
\newcommand{\rmint}{\mathrm{int}}

\newcommand{\calN}{\mathcal{N}}
\newcommand{\rmor}{\mathrm{or}}
\newcommand{\rmfor}{\mathrm{for}}
\newcommand{\rmtot}{\mathrm{tot}}
\newcommand{\rmc}{\mathrm{c}}

\newcommand{\rmg}{\mathrm{g}}
\newcommand{\rmwhere}{\mathrm{where}}

\newcommand{\rmwith}{\mathrm{with}}
\newcommand{\rmand}{\mathrm{and}}

\begin{document}
\title[Time-dependent pointer states of the generalized spin-boson model and etc.]{Time-dependent pointer states of the generalized spin-boson model and consequences regarding the decoherence of the central system}
\author{Hoofar Daneshvar and G W F Drake}
\address{Department of Physics, University of Windsor, Windsor ON, N9B 3P4, Canada}
\ead{hoofar@uwindsor.ca and GDrake@uwindsor.ca}

\begin{abstract}
We consider a spin-boson Hamiltonian which is generalized such that the Hamiltonians for the system ($\hat{H}_{\cal S}$) and the interaction with the environment ($\hat{H}_{\rm int}$) do not commute with each other. Considering a single-mode quantized field in exact resonance with the tunneling matrix element of the system, we obtain the time-evolution operator for our model. Using our time-evolution operator we calculate the time-dependent pointer states of the system and the environment (which are characterized by their ability not to entangle with each other) for the case that the environment initially is prepared in the coherent state. We show that our solution for the pointer states of the system and the environment is valid over a length of time which is proportional to $\bar{n}$, the average number of bosons in the environment. We also obtain a closed form for the offdiagonal element of the reduced density matrix of the system and study the decoherence of the central system in our model. We show that for the case that the system initially is prepared in one of its pointer states, the offdiagonal element of the reduced density matrix of the system will be a \emph{sinusoidal function} with a slow decaying envelope which is characterized by a decay time proportional to $\bar{n}$; while it will experience a much faster decoherence, when the system initially is \emph{not} prepared in one of its initial pointer states.
\end{abstract}


\section{Introduction}

This is the second paper in the series of papers where we discuss pointer states of measurement and their evaluation. We refer to this paper as paper $\sf II$ in this series of papers.
\textcolor[rgb]{0.00,0.00,0.00}{\subsection{Foreword}}
\textcolor[rgb]{0.00,0.00,0.00}{The pointer states of a system are defined as those states of the system characterized by their property of not entangling with states of another system \cite{Schlosshauer1,Zurek1,Zurek2}. This condition is commonly referred to as the \emph{stability
criterion} for the selection of pointer states \cite{Schlosshauer1,Zurek1}.}

\textcolor[rgb]{0.00,0.00,0.00}{In paper $\sf I$ \cite{paper1} we discussed the pointer states of measurement and presented a general method for obtaining the pointer states of the system and the environment for a given total Hamiltonian defining the system-environment model. As we elaborately described in paper $\sf I$, generally we should distinguish between the pointer states of a system and the preferred basis of measurement. We explicitly proved that the pointer states of a system generally are time-dependent and a preferred basis of measurement does not exist, unless under some specific conditions (discussed there in paper $\sf I$) that the pointer states of measurement become time-independent. Moreover, time-dependent pointer states necessarily are not orthogonal amongst themselves at all times. Therefore, they cannot be considered as eigenstates of a Hermitian operator at all times.}

\textcolor[rgb]{0.00,0.00,0.00}{In paper $\sf I$ we also used our method in order to rederive the time-dependent pointer states of the system and the environment for the Jaynes-Cummings model (JCM) of quantum optics \cite{paper1}; verifying the previous results for the JCM \cite{Gea-Banacloche,Gea-Banacloche2,Knight}.}

In this paper we study a spin-boson model \footnote{For a serious review and analysis of spin-boson models in different regimes the interested reader can refer to the seminal article by Leggett \emph{et al.} \cite{Leggett} or the book by Weiss \cite{Weiss}. Also a brief while very useful review of the model can be found in chapter 5 of Schlosshauer's book \cite{Schlosshauer1}.} which is defined through the following total Hamiltonian
\begin{equation}\label{1}
\hat{H}=-\frac{1}{2}\Delta_{0}\hat{\sigma}_{x}+\omega\hat{a}^{\dag}\hat{a}+\hat{\sigma}_{z}\otimes(\rmg \hat{a}^{\dag}+\rmg^{\ast}\hat{a}).
\end{equation}
This model basically is composed of a central spin-half particle (or other two-level system) surrounded by an environment of $N$ bosonic particles such as photons. In the Hamiltonian of the spin-boson model, represented by equation (\ref{1}), we have considered an intrinsic tunneling contribution in the self-Hamiltonian of the system (proportional to the $\hat{\sigma}_{x}$ Pauli matrix), which can induce intrinsic transitions between the upper and lower states of the central system. Here $\Delta_{0}$ is the so-called tunneling matrix element. Also, it is assumed that the asymmetry energy in the self-Hamiltonian of the central system is negligible. Therefore, here we do not consider a contribution proportional to the $\hat{\sigma}_{z}$ Pauli matrix in the self-Hamiltonian of the system \footnote{However, one can easily verify that modifications due to such contribution can be done quite easily, as such a term commutes with the interaction between the system and the environment, represented by the third term in equation (\ref{1}).}. The second term in equation (\ref{1}) represents the self-Hamiltonian of the electromagnetic field; where we have considered a single-mode quantized field with the frequency of $\omega$ for the environment. Also the third term, with $\rmg$ as the spin-field coupling constant,
represents the interaction between the central spin-half particle and a single-mode quantized field; which in fact is the quantized form of the famous
$-\bmu.\textbf{B}$
Hamiltonian due to the interaction between a particle of magnetic dipole-moment $\bmu$ and a magnetic field $\textbf{B}$.

\textcolor[rgb]{0.00,0.00,0.00}{In the Hamiltonian of equation (\ref{1}), if we switch the $\hat{\sigma}_{x}$ and $\hat{\sigma}_{z}$ operators and apply the rotating-wave approximation, we would obtain a Hamiltonian which would precisely look like the Hamiltonian of the Jaynes-Cummings model (JCM). However, the Hamiltonian that would be obtained in this way, will \emph{not} describe the Jaynes-Cummings model of quantum optics. The above point is because of the fact that when considering our model, represented by the Hamiltonian of equation (\ref{1}), a very special meaning is attributed to the eigenstates of the $\hat{\sigma}_{z}$ operator; they are the upper and lower states of the two-level system. In fact, this attribution of the upper and lower states of the two-level system to the eigenstates of the $\hat{\sigma}_{z}$ operator will be considered everywhere in our calculations; as is also considered in calculations for the JCM \cite{Scully,Drake}. Therefore, as we will see in the following section, when considering the rotating-wave-approximation with the exact resonance condition the Hamiltonian of equation (\ref{1}) will lead to a time-evolution operator that is very different from the one that we know for the JCM. As a result, as we will see, there will be clear differences between the physics which arises from our spin-boson model and the one that we know from the JCM.}

In the paper by Leggett \emph{et al.} \cite{Leggett} they considered a general form of the Hamiltonian given by equation (\ref{1}), where the environment can be represented by a spectral density \textbf{\emph{J}}$(\omega)$ (rather than considering a \emph{single-mode} quantized field). They used the ``influence-functional" method of Feynman and Vernon \cite{Feynman} to obtain general expressions for $P(t)\equiv\langle\hat{\sigma}_{z}(t)\rangle$ in the form of a power series in $\Delta$. However, the general expressions they obtained for $P(t)$ in terms of the spectral density function \textbf{\emph{J}}$(\omega)$ were exceedingly cumbersome to calculate in most regimes. So, they assumed \textcolor[rgb]{0.00,0.00,0.00}{a nonresonance regime, which was quite different from the resonance regime that we will consider here in this article} \cite{Leggett}.

The model represented by the Hamiltonian of equation (\ref{1}) can also be studied in the framework of the Born-Markov approximation in order to obtain an approximate master equation for the evolution of the reduced density matrix of the system \cite{Weiss}. The master equations obtained in this way are valid only in certain regimes; and moreover, one often may need to resort to numerical computation in order to be able to solve them. However, the main purpose of this paper is (1) to obtain the time-dependent pointer states of the system and the environment, as well as expressions for the evolution of the reduced density matrix of the system in the exact resonance regime and for an environment initially prepared in the coherent state; and (2) to obtain approximate expressions \emph{in closed form} for the evolution of the off-diagonal elements of the reduced density matrix (for the case that the environment initially is prepared in a coherent state with a large average number of photons), which can be used \textcolor[rgb]{0.00,0.00,0.00}{in order to} study the decoherence of the central system \emph{in an analytical way}.

\textcolor[rgb]{0.00,0.00,0.00}{\subsection{Review of our method for calculation of pointer states}}

\textcolor[rgb]{0.00,0.00,0.00}{Our goal is to find the pointer states of the system, as well as their corresponding pointer states from the environment for an arbitrary total Hamiltonian defining the system-environment model. In order to find pointer states, we look for system states
$|s_{i}(t)\rangle$ such that the composite system-environment state, when starting from a product state
$|s_{i}(t_{0})\rangle|E_{0}\rangle$ at $t=0$, remains in the product form
$|s_{i}(t)\rangle|E_{i}(t)\rangle$ at all subsequent times $t>0$; i.e.\ pointer states must satisfy the aforementioned stability criterion.}

Now consider a two-state system $\cal S$ with two arbitrary basis states $|a\rangle$ and $|b\rangle$, initially prepared in the state
\begin{equation}\label{240}
|\psi^{\cal S}(t_{0})\rangle=\alpha|a\rangle+\beta|b\rangle \quad \rmwith \quad |\alpha|^{2}+|\beta|^{2}=1;
\end{equation}
and an environment initially prepared in the state
\begin{equation}\label{250}
|\Phi^{\cal E}(t_{0})\rangle=\sum_{n=0}^{\infty}c_{n}|\varphi_{n}\rangle,
\end{equation}
where $\{|\varphi_{n}\rangle\}$'s are a complete set of basis states for the environment. \textcolor[rgb]{0.00,0.00,0.00}{For the two-state system with the two basis states $|a\rangle$ and $|b\rangle$ we can consider a time evolution operator for the global state of the system and the environment given by
\begin{equation}\label{270}
\hat{U}_{\rm tot}(t)=\hat{\cal E}_{1}|a\rangle\langle a|+\hat{\cal E}_{2}|a\rangle\langle b|
+\hat{\cal E}_{3}|b\rangle\langle a|+\hat{\cal E}_{4}|b\rangle\langle b|;
\end{equation}}
where the $\hat{\cal E}_{i}$'s are some generally \emph{time-dependent} operators in the Hilbert space of the environment, and depend on the total Hamiltonian defining the system-environment model.

Using equations (2) to (4) we can write the global state of the system and the environment as
\begin{eqnarray}\label{3.29}
\nonumber |\psi^{\rm tot}(t)\rangle=\hat{U}_{\rm tot}(t).\ (\alpha|a\rangle+\beta|b\rangle)\otimes(\sum_{n=0}^{\infty}c_{n}|\varphi_{n}\rangle)\\
=\textbf{A}(t)\ |a\rangle+\textbf{B}(t)\ |b\rangle; \\ \nonumber \rmwith \quad
\textbf{A}(t)=\sum_{n=0}^{\infty}c_{n}\{\alpha\hat{\cal E}_{1}(t)+\beta\hat{\cal E}_{2}(t)\}\ |\varphi_{n}\rangle \\ \rmand \quad \nonumber
\textbf{B}(t)=\sum_{n=0}^{\infty}c_{n}\{\alpha\hat{\cal E}_{3}(t)+\beta\hat{\cal E}_{4}(t)\}\ |\varphi_{n}\rangle.
\end{eqnarray}
Now, for the global state of the system and the environment, given by equation (\ref{3.29}), we observe that \emph{if} for some initial states of the system and the environment the two vectors $\textbf{A}(t)$ and $\textbf{B}(t)$ of the Hilbert space of the environment turn out to be parallel with each other, i.e.\ if
\begin{equation}\label{3.30.1}
\textbf{A}(t)=G(t)\textbf{B}(t),
\end{equation}
with $G(t)$ as a scalar which generally may depend on time, then those initial states of the system and the environment will not entangle with each other, and hence they can represent the initial pointer states of the system and the environment. In fact, if for some initial states of the system and the environment the condition represented by equation (\ref{3.30.1}) is satisfied, the global state of the system and the environment (equation (\ref{3.29})) can be written in a product from as
\begin{equation}\label{3.31}
\nonumber |\psi^{\rm tot}(t)\rangle=\{G(t)|a\rangle+|b\rangle\}\times(\sum_{n=0}^{\infty}c_{n}\{\alpha\hat{\cal E}_{3}+\beta\hat{\cal E}_{4}\}\ |\varphi_{n}\rangle),
\end{equation}
in which case pointer states can be realized for the system and the environment given by
\begin{eqnarray}\label{3.32}
|\pm(t)\rangle=\calN\ \{G(t)|a\rangle+|b\rangle\} \qquad \rmand\\ \nonumber
|\Phi_{\pm}(t)\rangle=\calN^{-1}\textbf{B}(t)=\calN^{-1}\sum_{n=0}^{\infty}c_{n}\{\alpha\hat{\cal E}_{3}(t)+\beta\hat{\cal E}_{4}(t)\}|\varphi_{n}\rangle.
\end{eqnarray}
In the above equation we have represented the pointer states of the system by $|\pm(t)\rangle$ and those of the environment by $|\Phi_{\pm}(t)\rangle$. Also, $\cal N$ is the normalization factor for the pointer states of the system.

In general, an arbitrary choice of $c_{n}$'s will not necessarily yield an $\textbf{A}(t), \textbf{B}(t)$ pair that remain parallel for any choice of $\alpha$ and $\beta$; therefore an arbitrary choice of $c_{n}$'s does not necessarily correspond to a pointer state for the environment. Nonetheless, for a given set of $c_{n}$'s there may exist two sets of complex numbers for $\alpha$ and $\beta$ with two corresponding values for the scalar function $G(t)$ such that equation (\ref{3.30.1}) is satisfied. In the following example, we show that for an assumed set of $c_{n}$'s, values of $\alpha$ and $\beta$ exist such that equation (\ref{3.30.1}) is approximately satisfied in the limit of large number of degrees of freedom for the environment \footnote{In fact, by looking at equation (\ref{3.31}) we notice that in practice we can expect some states of the system and the environment to keep their individuality and not to entangle with each other \emph{even} if they can satisfy our condition (given by equation (\ref{3.30.1})) only in a fraction of the Hilbert space of the environment where the $c_{n}$ coefficients are not negligible. This of course will involve assuming some approximations in obtaining pointer states. However, as we will show in this paper, in the end we can define a measure for the degree of entanglement between the states of the system and the environment, which after its calculation for the pointer states which we obtain for our model we can know exactly in which regimes our pointer states are valid and will not entangle with the states of another system. For example, this way we will show in this paper that the pointer states which we will obtain for our spin-boson model for an environment initially prepared in the coherent state are valid (i.e.\ will not entangle with the states of any other subsystem throughout their evolution with time) up to times of the order $\bar{n}/\rmg$; where $\bar{n}$ is the average number of photons in the coherent state of the environment.}.

\section{Calculation of the time-evolution operator}

In order to calculate the time-evolution operator in the interaction picture for the Hamiltonian given by equation (\ref{1}), first we need to have the corresponding Hamiltonian in the interaction picture. It can \textcolor[rgb]{0.00,0.00,0.00}{easily be calculated as}
\begin{equation}\label{6}
\hat{H}_{\rmint}(t)=\{\hat{\sigma}_{z}\cos(\Delta_{0}t)-\hat{\sigma}_{y}\sin(\Delta_{0}t)\}\{\rmg\hat{a}^{\dag}e^{i\omega t}+\rmg^{\ast}\hat{a}e^{-i\omega t}\}.
\end{equation}
Here, the commutator of $\hat{H}_{\rmint}(t)$ and $\hat{H}_{\rmint}(t'\neq t)$, i.e.\ $[\hat{H}_{\rmint}(t),\hat{H}_{\rmint}(t'\neq t)]$ with $\hat{H}_{\rmint}(t)$ given by equation (\ref{6}), is not a function of a constant number. This in fact can make the evaluation of the time-evolution operator quite difficult \cite{Duan}.

In parallel with paper $\sf I$ we consider the general form given by equation (\ref{270}) for the evolution operator of the global spin-field system.
For such time-evolution operator in the interaction picture, which satisfies the Schr\"{o}dinger equation
$i\hbar\frac{\partial}{\partial t}\hat{U}(t)=\hat{H}_{\rm int}\hat{U}(t),$
we have
\begin{eqnarray}\label{9}
i\hbar \left(
         \begin{array}{cc}
           \dot{\hat{\cal E}_{1}} & \dot{\hat{\cal E}_{2}} \\
           \dot{\hat{\cal E}_{3}} & \dot{\hat{\cal E}_{4}} \\
         \end{array}
       \right)=\hat{H}_{\rmint}(t)\left(
                 \begin{array}{cc}
                   \hat{\cal E}_{1} & \hat{\cal E}_{2} \\
                   \hat{\cal E}_{3} & \hat{\cal E}_{4} \\
                 \end{array}
               \right)=\{\rmg\hat{a}^{\dag}e^{i\omega t}+\rmg^{\ast}\hat{a}e^{-i\omega t}\}\\
               \nonumber \times\ \left(
                 \begin{array}{cc}
                   \hat{\cal E}_{1}\cos(\Delta_{0}t)+i\hat{\cal E}_{3}\sin(\Delta_{0}t) & \hat{\cal E}_{2}\cos(\Delta_{0}t)+i\hat{\cal E}_{4}\sin(\Delta_{0}t) \\
                   -\hat{\cal E}_{3}\cos(\Delta_{0}t)-i\hat{\cal E}_{1}\sin(\Delta_{0}t) & -\hat{\cal E}_{4}\cos(\Delta_{0}t)-i\hat{\cal E}_{2}\sin(\Delta_{0}t) \\
                 \end{array}
               \right).
\end{eqnarray}

Now, we assume the transition matrix element $\Delta_{0}$ to be in resonance with the cavity eigenmode $\omega$ and we use the rotating-wave approximation (RWA) \cite{Scully,Eberly} (just as is assumed in the conventional Jaynes-Cummings model of quantum optics \cite{Scully}). So, by entering the resonance condition $\Delta_{0}=\omega$ and using the rotating wave approximation (i.e.\ disregarding the higher-frequency terms which contain $e^{\pm i(\omega+\Delta_{0})t}$) the above equation will simplify to the following set of four equations
\begin{eqnarray}\label{10}
\nonumber i\dot{\hat{\cal E}_{1}}=\frac{\rmg\hat{a}^{\dag}}{2}(\hat{\cal E}_{1}-\hat{\cal E}_{3})+\frac{\rmg^{\ast}\hat{a}}{2}(\hat{\cal E}_{1}+\hat{\cal E}_{3}),\\
\nonumber i\dot{\hat{\cal E}_{2}}=\frac{\rmg\hat{a}^{\dag}}{2}(\hat{\cal E}_{2}-\hat{\cal E}_{4})+\frac{\rmg^{\ast}\hat{a}}{2}(\hat{\cal E}_{2}+\hat{\cal E}_{4}),\\
i\dot{\hat{\cal E}_{3}}=\frac{\rmg\hat{a}^{\dag}}{2}(\hat{\cal E}_{1}-\hat{\cal E}_{3})-\frac{\rmg^{\ast}\hat{a}}{2}(\hat{\cal E}_{1}+\hat{\cal E}_{3}),\\
\nonumber i\dot{\hat{\cal E}_{4}}=\frac{\rmg\hat{a}^{\dag}}{2}(\hat{\cal E}_{2}-\hat{\cal E}_{4})-\frac{\rmg^{\ast}\hat{a}}{2}(\hat{\cal E}_{2}+\hat{\cal E}_{4}).
\end{eqnarray}

In order to solve the above set of coupled differential equations, we proceed as follows. First, we take a derivative with respect to time of the first equation. By replacing $\dot{\hat{\cal E}_{1}}$ and $\dot{\hat{\cal E}_{3}}$ from the first and the third equations in the resulting equation we find
\begin{equation}\label{11}
i\ddot{\hat{\cal E}_{1}}=\frac{-i|\rmg|^{2}}{2}\{(1+2\hat{N})\hat{\cal E}_{1}-\hat{\cal E}_{3}\},
\end{equation}
where $\hat{N}=\hat{a}^{\dag}\hat{a}$ is the number operator. Similarly, by doing the same procedure on the third equation for $\dot{\hat{\cal E}_{3}}$ we find
\begin{equation}\label{12}
i\ddot{\hat{\cal E}_{3}}=\frac{-i|\rmg|^{2}}{2}\{(1+2\hat{N})\hat{\cal E}_{3}-\hat{\cal E}_{1}\}.
\end{equation}
Next, we define $\hat{\cal E}_{++}$ and $\hat{\cal E}_{+-}$ as follows
\begin{equation}\label{13}
\hat{\cal E}_{++}=\hat{\cal E}_{1}+\hat{\cal E}_{3} \qquad \rmand \qquad \hat{\cal E}_{+-}=\hat{\cal E}_{1}-\hat{\cal E}_{3}.
\end{equation}
By adding and subtracting equations (\ref{11}) and (\ref{12}) we find
\begin{eqnarray}\label{14}
\nonumber \ddot{\hat{\cal E}}_{++}=-|\rmg|^{2}\hat{N}\hat{\cal E}_{++} \qquad \rmand \\
\ddot{\hat{\cal E}}_{+-}=-|\rmg|^{2}(\hat{N}+1)\ \hat{\cal E}_{+-}\ .
\end{eqnarray}
These equations for $\hat{\cal E}_{++}$ and $\hat{\cal E}_{+-}$ can simply be solved to find
\begin{eqnarray}\label{15}
\nonumber \hat{\cal E}_{++}=\sin(|\rmg|\sqrt{\hat{N}}\ t)\hat{A}+\cos(|\rmg|\sqrt{\hat{N}}\ t)\hat{B}\qquad \rmand\\
\hat{\cal E}_{+-}=\sin(|\rmg|\sqrt{\hat{N}+1}\ t)\hat{A'}+\cos(|\rmg|\sqrt{\hat{N}+1}\ t)\hat{B'},
\end{eqnarray}
where $\hat{A}$, $\hat{A'}$, $\hat{B}$ and $\hat{B'}$ are some time-independent \emph{operators} (rather than constant \emph{numbers}), which will be found from our initial conditions in the following paragraphs. Here we note that since these coefficients generally are some time-independent operators rather than constant numbers, and they do not necessarily commute with the number operator $\hat{N}$, we \emph{must} have them on the right-hand side of the $sin$ and $cos$ functions (rather than having them on the left-hand side). Only in this way the solutions in equation (\ref{15}) will satisfy equation (\ref{14}).

Now using equations (\ref{13}) and (\ref{15}), we can obtain the operators $\hat{\cal E}_{1}$ and $\hat{\cal E}_{3}$ as follows:
\begin{eqnarray}\label{16}
\nonumber \hat{\cal E}_{1}=\frac{1}{2}\{\sin(|\rmg|\sqrt{\hat{N}}\ t)\hat{A}+\cos(|\rmg|\sqrt{\hat{N}}\ t)\hat{B}\\+\sin(|\rmg|\sqrt{\hat{N}+1}\ t)\hat{A'}+\cos(|\rmg|\sqrt{\hat{N}+1}\ t)\hat{B'}\} \qquad \rmand \\
\nonumber \hat{\cal E}_{3}=\frac{1}{2}\{\sin(|\rmg|\sqrt{\hat{N}}\ t)\hat{A}+\cos(|\rmg|\sqrt{\hat{N}}\ t)\hat{B}\\-\sin(|\rmg|\sqrt{\hat{N}+1}\ t)\hat{A'}-\cos(|\rmg|\sqrt{\hat{N}+1}\ t)\hat{B'}\}.
\end{eqnarray}
In quite the same manner we can calculate $\hat{\cal E}_{2}$ and $\hat{\cal E}_{4}$ as follows
\begin{eqnarray}\label{18}
\nonumber \hat{\cal E}_{2}=\frac{1}{2}\{\sin(|\rmg|\sqrt{\hat{N}}\ t)\hat{C}+\cos(|\rmg|\sqrt{\hat{N}}\ t)\hat{D}\\+\sin(|\rmg|\sqrt{\hat{N}+1}\ t)\hat{C'}+\cos(|\rmg|\sqrt{\hat{N}+1}\ t)\hat{D'}\} \qquad \rmand \\
\nonumber \hat{\cal E}_{4}=\frac{1}{2}\{\sin(|\rmg|\sqrt{\hat{N}}\ t)\hat{C}+\cos(|\rmg|\sqrt{\hat{N}}\ t)\hat{D}\\-\sin(|\rmg|\sqrt{\hat{N}+1}\ t)\hat{C'}-\cos(|\rmg|\sqrt{\hat{N}+1}\ t)\hat{D'}\};
\end{eqnarray}
where $\hat{C}$, $\hat{C'}$, $\hat{D}$ and $\hat{D'}$ also, generally are some time-independent \emph{operators}, which will be determined from our original set of equations (\ref{10}) and the initial conditions on $\{\hat{\cal E}_{i}\}$'s.

In order to obtain the eight operator coefficients which appear in our expressions for $\{\hat{\cal E}_{i}\}$'s, first we note that the time-evolution operator, given by equation (\ref{270}), must satisfy the initial condition $\hat{U}(t=0)=\hat{I}_{\cal S}\otimes\hat{I}_{\cal E}$; where $\hat{I}_{\cal S}=|a\rangle\langle a|+|b\rangle\langle b|$ represents the identity operator in the Hilbert space of the system and $\hat{I}_{\cal E}$ is the identity operator in the Hilbert space of the environment. This means that we must have
\begin{equation}\label{20}
\hat{\cal E}_{1}(0)=\hat{\cal E}_{4}(0)=\hat{I}_{\cal E} \qquad \rmand \qquad \hat{\cal E}_{2}(0)=\hat{\cal E}_{3}(0)=0
\end{equation}
From the above initial conditions and equations (\ref{16}) to (21) we easily find four of the coefficients as follows
\begin{equation}\label{21}
\hat{B}=\hat{B'}=\hat{D}=\hat{I}_{\cal E} \qquad \rmand \qquad \hat{D'}=-\hat{I}_{\cal E}.
\end{equation}

In order to find $\hat{A}$ and $\hat{A'}$ we proceed as follows. First, we use equation (\ref{10}-a) to obtain $\hat{\cal E}_{3}$ as follows
\begin{equation}\label{22}
(\rmg^{\ast}\hat{a}-\rmg\hat{a}^{\dag})\hat{\cal E}_{3}=2i\dot{\hat{\cal E}_{1}}-(\rmg^{\ast}\hat{a}+\rmg\hat{a}^{\dag})\hat{\cal E}_{1}\ .
\end{equation}
Replacing $\hat{\cal E}_{1}$ and $\dot{\hat{\cal E}_{1}}$ from equation (\ref{16}) into the above equation, it reads
\begin{eqnarray}\label{23}
\nonumber (\rmg^{\ast}\hat{a}-\rmg\hat{a}^{\dag})\hat{\cal E}_{3}=i|\rmg|\sqrt{\hat{N}}\ \{\cos(|\rmg|\sqrt{\hat{N}}\ t)\hat{A}-\sin(|\rmg|\sqrt{\hat{N}}\ t)\}\\+i|\rmg|\sqrt{\hat{N}+1}\ \{\cos(|\rmg|\sqrt{\hat{N}+1}\ t)\hat{A'}-\sin(|\rmg|\sqrt{\hat{N}+1}\ t)\}\\ \nonumber
-(\frac{\rmg^{\ast}\hat{a}+\rmg\hat{a}^{\dag}}{2})\{\sin(|\rmg|\sqrt{\hat{N}}\ t)\hat{A}+\cos(|\rmg|\sqrt{\hat{N}}\ t)\\ \nonumber+\sin(|\rmg|\sqrt{\hat{N}+1}\ t)\hat{A'}+\cos(|\rmg|\sqrt{\hat{N}+1}\ t)\}.
\end{eqnarray}
At $t=0$ the above equation reduces to
\begin{equation}\label{24}
i|\rmg|\sqrt{\hat{N}}\ \hat{A}+i|\rmg|\sqrt{\hat{N}+1}\ \hat{A'}-(\rmg^{\ast}\hat{a}+\rmg\hat{a}^{\dag})=(\rmg^{\ast}\hat{a}-\rmg\hat{a}^{\dag})\ \hat{\cal E}_{3}(t=0)=0.
\end{equation}
Operating this last equation on $|n\rangle$ we have
\begin{equation}\label{25}
i|\rmg|\sqrt{\hat{N}}\ \hat{A}\ |n\rangle+i|\rmg|\sqrt{\hat{N}+1}\ \hat{A'}\ |n\rangle=\rmg^{\ast}\sqrt{n}\ |n-1\rangle+\rmg\sqrt{n+1}\ |n+1\rangle.
\end{equation}
Next, we use equation (\ref{10}-c) to obtain $\hat{\cal E}_{1}$ as follows
\begin{equation}\label{26}
(\rmg\hat{a}^{\dag}-\rmg^{\ast}\hat{a})\hat{\cal E}_{1}=2i\dot{\hat{\cal E}_{3}}+(\rmg^{\ast}\hat{a}+\rmg\hat{a}^{\dag})\hat{\cal E}_{3}\ .
\end{equation}
Replacing $\hat{\cal E}_{3}$ and $\dot{\hat{\cal E}_{3}}$ from equation (19) into the above equation, it reads
\begin{eqnarray}\label{27}
\nonumber (\rmg\hat{a}^{\dag}-\rmg^{\ast}\hat{a})\hat{\cal E}_{1}=i|\rmg|\sqrt{\hat{N}}\ \{\cos(|\rmg|\sqrt{\hat{N}}\ t)\hat{A}-\sin(|\rmg|\sqrt{\hat{N}}\ t)\}\\+i|\rmg|\sqrt{\hat{N}+1}\ \{-\cos(|\rmg|\sqrt{\hat{N}+1}\ t)\hat{A'}+\sin(|\rmg|\sqrt{\hat{N}+1}\ t)\}\\ \nonumber
+(\frac{\rmg^{\ast}\hat{a}+\rmg\hat{a}^{\dag}}{2})\{\sin(|\rmg|\sqrt{\hat{N}}\ t)\hat{A}+\cos(|\rmg|\sqrt{\hat{N}}\ t)\\ \nonumber-\sin(|\rmg|\sqrt{\hat{N}+1}\ t)\hat{A'}-\cos(|\rmg|\sqrt{\hat{N}+1}\ t)\}.
\end{eqnarray}
At $t=0$ the above equation reduces to
\begin{equation}\label{28}
i|\rmg|\sqrt{\hat{N}}\ \hat{A}-i|\rmg|\sqrt{\hat{N}+1}\ \hat{A'}=(\rmg\hat{a}^{\dag}-\rmg^{\ast}\hat{a})\ \hat{\cal E}_{1}(t=0)=(\rmg\hat{a}^{\dag}-\rmg^{\ast}\hat{a}).
\end{equation}
Operating this last equation on $|n\rangle$ we have
\begin{equation}\label{29}
i|\rmg|\sqrt{\hat{N}}\ \hat{A}\ |n\rangle-i|\rmg|\sqrt{\hat{N}+1}\ \hat{A'}\ |n\rangle=\rmg\sqrt{n+1}\ |n+1\rangle-\rmg^{\ast}\sqrt{n}\ |n-1\rangle.
\end{equation}
Finally, we use equations (\ref{25}) and (\ref{29}) to obtain the coefficients $\hat{A}$ and $\hat{A'}$. Assuming $\rmg$ to be real and then adding equations (\ref{25}) and (\ref{29}) we find
\begin{equation}\label{30}
\hat{A}\ |n\rangle=-i\sqrt{\frac{n+1}{\hat{N}}}\ |n+1\rangle=-i|n+1\rangle.
\end{equation}
Comparing the above equation to $\frac{-i}{\sqrt{\hat{N}}}\ \hat{a}^{\dag}\ |n\rangle=-i|n+1\rangle$ we find $\hat{A}$ as
\begin{equation}\label{31}
\hat{A}=\frac{-i}{\sqrt{\hat{N}}}\ \hat{a}^{\dag}.
\end{equation}
Similarly, by subtracting equation (\ref{29}) from equation (\ref{25}) to find
\begin{equation}\label{32}
\hat{A'}\ |n\rangle=-i\sqrt{\frac{n}{\hat{N}+1}}\ |n-1\rangle=-i|n-1\rangle
\end{equation}
and comparing the above equation to $\frac{-i}{\sqrt{\hat{N}+1}}\ \hat{a}\ |n\rangle=-i|n-1\rangle$ we find $\hat{A'}$ as
\begin{equation}\label{33}
\hat{A'}=\frac{-i}{\sqrt{\hat{N}+1}}\ \hat{a}\ .
\end{equation}
One should pay attention to the order that the operators appear in equations (\ref{31}) and (\ref{33}); as they are not commuting operators.

By doing exactly the same procedure on equations (\ref{10}-b) and (\ref{10}-d) we would find the operator coefficients $\hat{C}$ and $\hat{C'}$ as follows
\begin{equation}\label{34}
\hat{C}=\frac{i}{\sqrt{\hat{N}}}\ \hat{a}^{\dag}=-\hat{A} \qquad \rmand \qquad \hat{C'}=\frac{-i}{\sqrt{\hat{N}+1}}\ \hat{a}=\hat{A'}\ .
\end{equation}
Now that we found all the operator coefficients, we can replace them back in equations (\ref{16}) to (20) and write the $\{\hat{\cal E}_{i}\}$ in their final form
\begin{eqnarray}\label{35}
\nonumber \hat{\cal E}_{1}=\frac{1}{2}\{-i\sin(|\rmg|\sqrt{\hat{N}}\ t)\frac{1}{\sqrt{\hat{N}}}\ \hat{a}^{\dag}+\cos(|\rmg|\sqrt{\hat{N}}\ t)\\-i\sin(|\rmg|\sqrt{\hat{N}+1}\ t)\frac{1}{\sqrt{\hat{N}+1}}\ \hat{a}+\cos(|\rmg|\sqrt{\hat{N}+1}\ t)\}, \\
\nonumber \hat{\cal E}_{3}=\frac{1}{2}\{-i\sin(|\rmg|\sqrt{\hat{N}}\ t)\frac{1}{\sqrt{\hat{N}}}\ \hat{a}^{\dag}+\cos(|\rmg|\sqrt{\hat{N}}\ t)\\+i\sin(|\rmg|\sqrt{\hat{N}+1}\ t)\frac{1}{\sqrt{\hat{N}+1}}\ \hat{a}-\cos(|\rmg|\sqrt{\hat{N}+1}\ t)\},
\end{eqnarray}
\begin{eqnarray}\label{36}
\nonumber \hat{\cal E}_{2}=\frac{1}{2}\{i\sin(|\rmg|\sqrt{\hat{N}}\ t)\frac{1}{\sqrt{\hat{N}}}\ \hat{a}^{\dag}+\cos(|\rmg|\sqrt{\hat{N}}\ t)\\-i\sin(|\rmg|\sqrt{\hat{N}+1}\ t)\frac{1}{\sqrt{\hat{N}+1}}\ \hat{a}-\cos(|\rmg|\sqrt{\hat{N}+1}\ t)\} \qquad \rmand \\
\nonumber \hat{\cal E}_{4}=\frac{1}{2}\{i\sin(|\rmg|\sqrt{\hat{N}}\ t)\frac{1}{\sqrt{\hat{N}}}\ \hat{a}^{\dag}+\cos(|\rmg|\sqrt{\hat{N}}\ t)\\+i\sin(|\rmg|\sqrt{\hat{N}+1}\ t)\frac{1}{\sqrt{\hat{N}+1}}\ \hat{a}+\cos(|\rmg|\sqrt{\hat{N}+1}\ t)\}.
\end{eqnarray}
One can verify that the above set of operators satisfy the unitarity of the time-evolution operator $\hat{U}^{\dag}\hat{U}=\hat{U}\hat{U}^{\dag}=\hat{I}$.

\section{Calculation of the time-dependent pointer states of the system and the environment}
In order to obtain the time-dependent pointer states of the system and those of its environment for our model, represented by the Hamiltonian of equation (\ref{1}), we assume the field to be initially prepared in the coherent state
$|\nu\rangle$
\begin{equation}\label{48}
    |\Phi_{\rm
    field}(t_{0})\rangle=|\nu\rangle=\sum_{n=0}^{\infty}\rmc_{n}|n\rangle;
    \quad\rmwith \quad
    \rmc_{n}=\frac{\rme^{-\frac{1}{2}|\nu|^{2}}\nu^{n}}{\sqrt{n!}},
\end{equation}
where $|\nu|^{2}=\bar{n}$ is the average number of photons in the
coherent state, and $\nu=|\nu|e^{-i\varphi}$. In this section we will show that in the regime that we are considering (i.e.\ the exact-resonance \textcolor[rgb]{0.00,0.00,0.00}{with the RWA} regime) and for the environment initially prepared in the coherent state, in the limit of a large average number of photons $\bar{n}\rightarrow\infty$ we must have pointer states for the system (the central spin) given by
\begin{eqnarray}\label{49}
\nonumber |+(t)\rangle=-i\cos(\frac{\varphi}{2}+\frac{\rmg t}{4\sqrt{\bar{n}}})|a\rangle+\sin(\frac{\varphi}{2}+\frac{\rmg t}{4\sqrt{\bar{n}}})|b\rangle \qquad \rmand \\ |-(t)\rangle=i\sin(\frac{\varphi}{2}-\frac{\rmg t}{4\sqrt{\bar{n}}})|a\rangle+\cos(\frac{\varphi}{2}-\frac{\rmg t}{4\sqrt{\bar{n}}})|b\rangle,
\end{eqnarray}
where $|a\rangle$ and $|b\rangle$ are eigenstates of the $\hat{\sigma}_{z}$ Pauli matrix.

We make the usual assumption that there exists no correlations between the system and the environment at $t=0$. So, we consider the following initial state for the total composite system
\begin{equation}\label{50}
|\psi^{\rmtot}(t_{0})\rangle=(\alpha|a\rangle+\beta|b\rangle)\otimes\sum_{n=0}^{\infty}\rmc_{n}|n\rangle \qquad \rmwith \qquad |\alpha|^{2}+|\beta|^{2}=1.
\end{equation}
\textcolor[rgb]{0.00,0.00,0.00}{For pointer states the two vectors $\textbf{A}(t)$ and $\textbf{B}(t)$ of the Hilbert space of the environment must be parallel with each other. Therefore, for pointer states due to our condition, given by equation (\ref{3.30.1}), we must have:}
\begin{equation}\label{51}
\nonumber \sum_{n}c_{n}\{\alpha\hat{\cal E}_{1}+\beta\hat{\cal E}_{2}\}\ |\varphi_{n}\rangle=G(t)\times\sum_{n}c_{n}\{\alpha\hat{\cal E}_{3}+\beta\hat{\cal E}_{4}\}\ |\varphi_{n}\rangle.
\end{equation}
For the $\{\hat{\cal E}_{i}\}$, given by equations (36) through (39), we can write
\begin{equation}\label{52}
\hat{\cal E}_{i}|n\rangle=f_{i1}(n)|n+1\rangle+f_{i2}(n)|n\rangle+f_{i3}(n)|n-1\rangle,
\end{equation}
where $f_{ij}$'s (with $i=1,2,3,4$ and $j=1,2,3$) are given by
\begin{eqnarray}\label{53}
\nonumber f_{11}(n)=f_{31}(n)=-f_{21}(n)=-f_{41}(n)=\frac{-i}{2}\sin(\rmg t\sqrt{n+1})\equiv f_{1}(n),\\
\nonumber f_{12}(n)=f_{42}(n)=\frac{1}{2}(\cos(\rmg t\sqrt{n})+\cos(\rmg t\sqrt{n+1}))\equiv f_{2}(n),\\
f_{22}(n)=f_{32}(n)=\frac{1}{2}(\cos(\rmg t\sqrt{n})-\cos(\rmg t\sqrt{n+1}))\equiv f_{3}(n) \qquad \rmand\\
\nonumber f_{13}(n)=f_{23}(n)=-f_{33}(n)=-f_{43}(n)=\frac{-i}{2}\sin(\rmg t\sqrt{n})\equiv f'_{1}(n).
\end{eqnarray}
Using equation (\ref{52}) our condition, given by equation (\ref{51}), becomes
\begin{eqnarray}\label{54}
\nonumber \sum_{n=0}^{\infty}c_{n}\{\alpha(f_{11}(n)|n+1\rangle+f_{12}(n)|n\rangle+f_{13}(n)|n-1\rangle)\\ \nonumber +\beta(f_{21}(n)|n+1\rangle+
f_{22}(n)|n\rangle+f_{23}(n)|n-1\rangle)\}\\=G(t)\times \sum_{n=0}^{\infty}c_{n}\{\alpha(f_{31}(n)|n+1\rangle+f_{32}(n)|n\rangle+f_{33}(n)|n-1\rangle)\\ \nonumber +\beta(f_{41}(n)|n+1\rangle+f_{42}(n)|n\rangle+f_{43}(n)|n-1\rangle)\}.
\end{eqnarray}

\textcolor[rgb]{0.00,0.00,0.00}{Now}, since the number states $\{|n\rangle\}$ are a complete set of basis states for the environment, for the initial pointer states we can open the summations in equation (\ref{54}) and equalize terms from the two sides of this equation which correspond to the same number state $|n\rangle$. After using equation (\ref{53}), equation (\ref{54}) would simplify as
\textcolor[rgb]{0.00,0.00,0.00}{\begin{eqnarray}\label{56}
\fl \nonumber G(t)=\frac{(\alpha-\beta)c_{n}f_{1}(n)+\alpha c_{n+1}f_{2}(n+1)+\beta c_{n+1}f_{3}(n+1)+(\alpha+\beta)c_{n+2}f'_{1}(n+2)}{(\alpha-\beta)c_{n}f_{1}(n)+\alpha c_{n+1}f_{3}(n+1)+\beta c_{n+1}f_{2}(n+1)-(\alpha+\beta)c_{n+2}f'_{1}(n+2)};
\\ for\ all\ n.
\end{eqnarray}}
The above result for $G(t)$, which generally depends on $n$, would contradict our initial assumption of \textcolor[rgb]{0.00,0.00,0.00}{the two vectors $\textbf{A}(t)$ and $\textbf{B}(t)$ being parallel to each other}, \emph{unless} we can find certain initial states for the system for which $G(t)$ turns out to become independent of $n$; since as we discussed, for pointer states all components of the vector $\textbf{A}$ (${A_{n}}^{'}s$) must be related to their corresponding components from $\textbf{B}$ (${B_{n}}^{'}s$) through the \emph{same} scalar factor $G$ (see equation (6)). So now we should seek for those particular initial states of the system which can make $G(t)$ independent of the index $n$ of the states of the environment.

For an initial coherent field (equation (\ref{48})) we have: \\ $c_{n+1}=c_{n}e^{-i\varphi}\sqrt{\frac{\bar{n}}{n+1}}$ and $c_{n+2}=c_{n}e^{-2i\varphi}\frac{\bar{n}}{\sqrt{(n+1)(n+2)}}$.
Moreover, in the limit of a large average number of photons $\bar{n}\rightarrow\infty$ we can replace the factors $\sqrt{\frac{\bar{n}}{n+1}}$ and $\sqrt{\frac{\bar{n}}{n+2}}$ by unity\footnote{This approximation has been used by Gea-Banacloche and other people \cite{Gea-Banacloche,Gea-Banacloche2} in the study of the Jaynes-Cummings model of quantum optics. In fact, for the Jaynes-Cummings model it has been shown that an average number of photons only as large as twenty is enough to make this assumption a good approximation \cite{Gea-Banacloche}.};
since for $\bar{n}\rightarrow\infty$ the Poisson distribution of the coherent field is extremely sharp (with $\bar{n}$ at the center) and hence, for $\bar{n}\rightarrow\infty$ and $n\approx\bar{n}$ we have $\sqrt{\frac{\bar{n}}{n+1}}\approx1$ and $\sqrt{\frac{\bar{n}}{n+2}}\approx1$, while for $n$ being far from $\bar{n}$ the $c_{n}$ coefficient is negligible. So, the corresponding terms (of $n$ being far from $\bar{n}$) do not have any contribution in the summations of equation (\ref{54}). As a result, equation (\ref{56}) for $G(t)$ can be further simplified to
\begin{eqnarray}\label{57}
\fl G(t)=\frac{(\alpha-\beta)f_{1}(n)+\alpha e^{-i\varphi}f_{2}(n+1)+\beta e^{-i\varphi}f_{3}(n+1)+(\alpha+\beta)e^{-2i\varphi}f'_{1}(n+2)}{(\alpha-\beta)f_{1}(n)+\alpha e^{-i\varphi}f_{3}(n+1)+\beta e^{-i\varphi}f_{2}(n+1)-(\alpha+\beta)e^{-2i\varphi}f'_{1}(n+2)}.
\end{eqnarray}
Replacing the $f_{i}$ functions from equation (\ref{53}), the above equation reads
\begin{eqnarray}\label{58}
\nonumber G(t)=\{\textcolor[rgb]{0.98,0.00,0.00}{(\alpha+\beta)e^{-i\varphi}}\cos(\rmg t\sqrt{n+1})-i\textcolor[rgb]{0.98,0.00,0.00}{(\alpha-\beta)}\sin(\rmg t\sqrt{n+1}) \\ \nonumber+\textcolor[rgb]{0.00,0.00,1.00}{(\alpha-\beta)e^{-i\varphi}}\cos(\rmg t\sqrt{n+2})-i\textcolor[rgb]{0.00,0.00,1.00}{(\alpha+\beta)e^{-2i\varphi}}\sin(\rmg t\sqrt{n+2})\} \\ \quad \div\ \quad \{\textcolor[rgb]{0.98,0.00,0.00}{(\alpha+\beta)e^{-i\varphi}}\cos(\rmg t\sqrt{n+1})-i\textcolor[rgb]{0.98,0.00,0.00}{(\alpha-\beta)}\sin(\rmg t\sqrt{n+1})\nonumber \\ +\textcolor[rgb]{0.00,0.00,1.00}{(\beta-\alpha)e^{-i\varphi}}\cos(\rmg t\sqrt{n+2})+i\textcolor[rgb]{0.00,0.00,1.00}{(\alpha+\beta)e^{-2i\varphi}}\sin(\rmg t\sqrt{n+2})\}.
\end{eqnarray}

In order to obtain the pointer states of the system, we should look for the probable initial states of the system (represented by the coefficients $\alpha$ and $\beta$ in equation (\ref{50})) which can make the expression (in the above equation) for $G(t)$ \emph{independent} of the index $n$ of the states of the environment. On the other hand, by looking at equation (\ref{58}) we realize that if $\alpha-\beta=\pm(\alpha+\beta)e^{-i\varphi}$ the expression for $G(t)$ will be considerably simplified. In what follows we show that for $\alpha-\beta=\pm(\alpha+\beta)e^{-i\varphi}$, which is equivalent to the initial conditions for the state of the system given by
\begin{eqnarray}\label{59}
\nonumber
\alpha_{+}=-i\cos(\varphi/2)\quad \rmand \quad \beta_{+}=\sin(\varphi/2)\ \ \rm(for\ the\ plus\ sign)\quad\ \rmor \\
\alpha_{-}=i\sin(\varphi/2)\qquad \rmand \qquad \beta_{-}=\cos(\varphi/2)\ \ \rm(for\ the\ minus\ sign),
\end{eqnarray}
$G(t)$ of equation (\ref{58}) will be independent of the states of the environment; provided we have a large average number of photons in the field $\bar{n}\rightarrow\infty$. Therefore, the initial conditions of equation (\ref{59}) correspond to the initial states of the system which do not entangle with the states of the environment. After that, we will obtain the time evolution of these initial pointer states and followed by that we obtain the corresponding pointer states of the environment.

For $\alpha-\beta=\pm(\alpha+\beta)e^{-i\varphi}$ the expression in equation (\ref{58}) for $G(t)$ simplifies to
\begin{equation}\label{60}
G(t)=\frac{e^{\mp i\rmg t\sqrt{n+1}}\pm e^{-i\varphi}\ e^{\mp i\rmg t\sqrt{n+2}}}{e^{\mp i\rmg t\sqrt{n+1}}\mp e^{-i\varphi}\ e^{\mp i\rmg t\sqrt{n+2}}}.
\end{equation}
The above expression can be written as
\begin{eqnarray}\label{61}
\fl \nonumber G(t)=\frac{ \{e^{-i\varphi/2}\ e^{\mp\frac{i\rmg t}{2}(\sqrt{n+1}+\sqrt{n+2})}\} \{e^{i\varphi/2}\ e^{\pm\frac{i\rmg t}{2}(\sqrt{n+2}-\sqrt{n+1})}\pm e^{-i\varphi/2}\ e^{\pm\frac{i\rmg t}{2}(\sqrt{n+1}-\sqrt{n+2})}\} }   {\{e^{-i\varphi/2}\ e^{\mp\frac{i\rmg t}{2}(\sqrt{n+1}+\sqrt{n+2})}\} \{e^{i\varphi/2}\ e^{\pm\frac{i\rmg t}{2}(\sqrt{n+2}-\sqrt{n+1})}\mp e^{-i\varphi/2}\ e^{\pm\frac{i\rmg t}{2}(\sqrt{n+1}-\sqrt{n+2})}\} }\\ =\frac{ \{e^{i\varphi/2}\ e^{\pm\frac{i\rmg t}{2}(\sqrt{n+2}-\sqrt{n+1})}\pm e^{-i\varphi/2}\ e^{\pm\frac{i\rmg t}{2}(\sqrt{n+1}-\sqrt{n+2})}\} }   { \{e^{i\varphi/2}\ e^{\pm\frac{i\rmg t}{2}(\sqrt{n+2}-\sqrt{n+1})}\mp e^{-i\varphi/2}\ e^{\pm\frac{i\rmg t}{2}(\sqrt{n+1}-\sqrt{n+2})}\} }.
\end{eqnarray}
From the Taylor series expansion of $\sqrt{n+2}-\sqrt{n+1}$ about $\bar{n}$ (the average number of photons in the environment), given by
\begin{equation}\label{61.1}
\sqrt{n+2}-\sqrt{n+1}=\frac{1}{2\sqrt{\bar{n}}}-\frac{(n+1-\bar{n})}{4\bar{n}^{3/2}}+ ...\ ,
\end{equation}
we notice that in the limit of a very large average number of photons we can replace $\sqrt{n+2}-\sqrt{n+1}$ by $\frac{1}{2\sqrt{\bar{n}}}$ \cite{Gea-Banacloche}; since for very large $\bar{n}$ all terms containing the index $n$, which appear after the first term, are negligible; and \emph{the series is convergent}. 
So, in the classical limit of $\bar{n}\rightarrow\infty$ we can rewrite equation (\ref{61}) for $G(t)$ as
 \begin{eqnarray}\label{62}
\nonumber G(t)=\frac{ \{e^{i\varphi/2}\ e^{\pm\frac{i\rmg t}{4\sqrt{\bar{n}}}}\pm e^{-i\varphi/2}\ e^{\mp\frac{i\rmg t}{4\sqrt{\bar{n}}}}\} }   { \{e^{i\varphi/2}\ e^{\pm\frac{i\rmg t}{4\sqrt{\bar{n}}}}\mp e^{-i\varphi/2}\ e^{\mp\frac{i\rmg t}{4\sqrt{\bar{n}}}}\} } \\ \nonumber
=-i\cot(\frac{\varphi}{2}+\frac{\rmg t}{4\sqrt{\bar{n}}}) \qquad \rm{for\ the\ first\ sign}\ \ \\ \rmor \qquad=i\tan(\frac{\varphi}{2}-\frac{\rmg t}{4\sqrt{\bar{n}}}) \qquad \rm{for\ the\ second\ sign};
\end{eqnarray}
which clearly is independent of the index $n$ of the states of the environment.

In appendix A by calculating the degree of entanglement between the states of the system and the environment for the pointer states which will be obtained from the above result, we will show that this result is valid over a length of time which is proportional to $\bar{n}$, the average number of photons in the field.

The result of equation (\ref{62})) simply means that for the initial states of the system given by
\begin{equation}\label{63}
\fl |+(t_{0})\rangle=-i\cos(\varphi/2)|a\rangle+\sin(\varphi/2)|b\rangle \quad \rmand \quad |-(t_{0})\rangle=i\sin(\varphi/2)|a\rangle+\cos(\varphi/2)|b\rangle
\end{equation}
the states of the system and the environment will not entangle with each other. Moreover, using equation (\ref{3.32}) which gives us the general time evolution of the pointer states of the system; and $G(t)$ of equation (\ref{62}) (which is independent of the index $n$ of the states of the environment) we can find the time evolution of the pointer states of the system as follows
\begin{eqnarray}\label{64}
\nonumber |+(t)\rangle=\calN_{+}\ \{-i\cot(\frac{\varphi}{2}+\frac{\rmg t}{4\sqrt{\bar{n}}})|a\rangle+|b\rangle\} \quad \rmand \\
|-(t)\rangle=\calN_{-}\ \{i\tan(\frac{\varphi}{2}-\frac{\rmg t}{4\sqrt{\bar{n}}})|a\rangle+|b\rangle\};
\end{eqnarray}
where $\calN_{+}$ and $\calN_{-}$ are the normalization factors for the $|+(t)\rangle$ and $|-(t)\rangle$ states respectively. It is easy to verify that
\begin{equation}\label{65}
\fl \qquad \quad \calN_{+}=\sin(\theta_{+}(t)) \quad \rmand \quad \calN_{-}=\cos(\theta_{-}(t)), \quad \rmwhere \quad \theta_{\pm}(t)=\frac{\varphi}{2}\pm\frac{\rmg t}{4\sqrt{\bar{n}}}.
\end{equation}
So, we can rewrite equation (\ref{64}) as
\begin{eqnarray}\label{66}
\nonumber |+(t)\rangle=-i\cos(\frac{\varphi}{2}+\frac{\rmg t}{4\sqrt{\bar{n}}})|a\rangle+\sin(\frac{\varphi}{2}+\frac{\rmg t}{4\sqrt{\bar{n}}})|b\rangle \qquad \rmand \\ |-(t)\rangle=i\sin(\frac{\varphi}{2}-\frac{\rmg t}{4\sqrt{\bar{n}}})|a\rangle+\cos(\frac{\varphi}{2}-\frac{\rmg t}{4\sqrt{\bar{n}}})|b\rangle,
\end{eqnarray}
which is the same as equation (\ref{49}); Q.E.D.

Next, we obtain the corresponding pointer states of the environment. Using equations (\ref{3.32}) and (\ref{52}) we have
\begin{eqnarray}\label{67}
\nonumber |\Phi_{\pm}(t)\rangle=\calN_{\pm}^{-1}\sum_{n=0}^{\infty}c_{n}\{\alpha_{\pm}\hat{\cal E}_{3}+\beta_{\pm}\hat{\cal E}_{4}\}\ |\varphi_{n}\rangle \\ =
\calN_{\pm}^{-1}\sum_{n=0}^{\infty}c_{n}\{\alpha_{\pm}\ [f_{31}(n)|n+1\rangle+f_{32}(n)|n\rangle+f_{33}(n)|n-1\rangle]\\ \nonumber +\beta_{\pm}\  [f_{41}(n)|n+1\rangle+f_{42}(n)|n\rangle+f_{43}(n)|n-1\rangle]\},
\end{eqnarray}
where in the above equation $\alpha_{\pm}$ and $\beta_{\pm}$ are those of the initial pointer states of the system given by equation (\ref{59}). Let us first obtain $|\Phi_{+}(t)\rangle$; i.e.\ the pointer state of the environment corresponding to the $|+(t)\rangle$ state. Replacing the $f_{ij}$ functions from equation (\ref{53}), $\alpha_{+}$ and $\beta_{+}$ from equation (\ref{59}) and $\calN_{+}$ from equation (\ref{65}) we have
\begin{eqnarray}\label{68}
\nonumber |\Phi_{+}(t)\rangle=\sum_{n=0}^{\infty}\frac{c_{n}}{2\sin(\theta_{+}(t))}\{-\sin(\rmg t\sqrt{n+1})\ e^{-i\varphi/2}\ |n+1\rangle \\ \nonumber +[-i\cos(\rmg t\sqrt{n})\ e^{i\varphi/2}+i\cos(\rmg t\sqrt{n+1})\ e^{-i\varphi/2}\ ]\ |n\rangle \\+\sin(\rmg t\sqrt{n})\ e^{i\varphi/2}\ |n-1\rangle\}.
\end{eqnarray}
Using $c_{n\pm1}\approx c_{n}e^{\mp i\varphi}$ for the coherent field and in the limit of $\bar{n}\rightarrow\infty$, the above relation can be written as
\begin{eqnarray}\label{69}
\nonumber |\Phi_{+}(t)\rangle=\sum_{n=0}^{\infty}\frac{c_{n}}{2i\sin(\theta_{+}(t))}\{-i\sin(\rmg t\sqrt{n})\ e^{i\varphi/2}\ \\ \nonumber +[\cos(\rmg t\sqrt{n})\ e^{i\varphi/2}-\cos(\rmg t\sqrt{n+1})\ e^{-i\varphi/2}\ ]\ \\ \nonumber +i\sin(\rmg t\sqrt{n+1})\ e^{-i\varphi/2}\ \}\ |n\rangle \\ =\sum_{n=0}^{\infty}\frac{c_{n}}{2i\sin(\theta_{+}(t))}\ \{e^{i(\varphi/2-\rmg t\sqrt{n})}-e^{-i(\varphi/2+\rmg t\sqrt{n+1})}\}\ |n\rangle,
\end{eqnarray}
which can easily be simplified into the following final result for $|\Phi_{+}(t)\rangle$
\begin{equation}\label{70}
|\Phi_{+}(t)\rangle=\sum_{n=0}^{\infty}c_{n}\ e^{\frac{-i\rmg t}{2}(\sqrt{n+1}+\sqrt{n})}\ |n\rangle.
\end{equation}
Following exactly the same procedure one can also find $|\Phi_{-}(t)\rangle$ as follows
\begin{equation}\label{71}
|\Phi_{-}(t)\rangle=\sum_{n=0}^{\infty}c_{n}\ e^{\frac{i\rmg t}{2}(\sqrt{n+1}+\sqrt{n})}\ |n\rangle.
\end{equation}

We should also mention that the pointer states of the system at $t=t_{0}$ (equation (\ref{63})) are orthonormal and hence, they form a complete basis set for the state of the system. Therefore, the evolution of \emph{any} initial state of the system $|\psi_{\cal S}(t_{0})\rangle=\alpha'\ |+(t_{0})\rangle+\beta'\ |-(t_{0})\rangle$ in contact with an initial coherent field $|\nu\rangle$ can be expressed as a linear combination of the evolution of $|+(t_{0})\rangle|\nu\rangle$ and $|-(t_{0})\rangle|\nu\rangle$, and in the following diagonal form:
\begin{eqnarray}\label{73}
    \fl (\alpha'\ |+(t_{0})\rangle+\beta'\ |-(t_{0})\rangle)\ |\nu\rangle\rightarrow
    \alpha'\ |+(t)\rangle\ |\Phi_{+}(t)\rangle+\beta'\ |-(t)\rangle\ |\Phi_{-}(t)\rangle.
\end{eqnarray}

\section{State preparation at specific times}
One interesting feature of the pointer states of the system, given by equation (\ref{66}), is that at specific times they coincide with each other. In fact, by looking at equation (\ref{64}) we notice that at those times for which $\tan(\frac{\varphi}{2}-\frac{\rmg t}{4\sqrt{\bar{n}}})=-\cot(\frac{\varphi}{2}+\frac{\rmg t}{4\sqrt{\bar{n}}})$ the $|\pm(t)\rangle$ states are equal to each other. One can easily verify that at $t_{1}=(4n+1)\pi\sqrt{\bar{n}}/\rmg \ (\rmwith\ n=0,1,2,...)$ both of the $|\pm(t)\rangle$ states will be in the common state given by
\begin{equation}\label{74}
|\pm(t_{1})\rangle=i\sin(\frac{\varphi}{2}-\frac{\pi}{4})|a\rangle+\cos(\frac{\varphi}{2}-\frac{\pi}{4})|b\rangle;
\end{equation}
while at $t_{2}=(4n-1)\pi\sqrt{\bar{n}}/\rmg \ (\rmwith\ n=1,2,...)$ the $|\pm(t)\rangle$ states will be in the common state given by
\begin{equation}\label{75}
|\pm(t_{2})\rangle=i\sin(\frac{\varphi}{2}+\frac{\pi}{4})|a\rangle+\cos(\frac{\varphi}{2}+\frac{\pi}{4})|b\rangle.
\end{equation}

The state preparation at these specific times basically means that whatever is the initial state of the system, at these specific times the states of the system and the environment are not entangled to each other and the system can be represented by a well-defined state of its own (see equation (\ref{73})). Moreover, as we see, these specific states clearly depend on the phase $\varphi$ of the initial state of the coherent field.
The same kind of phenomenon was also discovered in the simpler Jaynes-Cummings model of quantum optics by Gea-Banacloche \cite{Gea-Banacloche} in 1991.

\section{Consequences regarding the decoherence of the central spin}
In this section we will use the pointer states of the system and the environment, which we already obtained in section 3, in order to \textcolor[rgb]{0.00,0.00,0.00}{obtain a \emph{closed} form for the coherences of the reduced density matrix of the system}. However, prior to that let us use the time-evolution operator, which we already obtained in section 2, to calculate the offdiagonal element of the reduced density matrix of the system \textcolor[rgb]{0.00,0.00,0.00}{in a more} precise form.

\textcolor[rgb]{0.00,0.00,0.00}{\subsection{General expressions for the evolution of the state of the total composite system and the reduced density matrix of the system}}

Using equations (\ref{3.29}) and (\ref{52}) to obtain $|\psi_{\rmtot}(t)\rangle$, we can write
\begin{eqnarray}\label{76}
\fl \nonumber |\psi_{\rmtot}(t)\rangle=\sum_{n=0}^{\infty}c_{n}\{(\alpha f_{11}(n)+\beta f_{21}(n))|a,n+1\rangle+(\alpha f_{12}(n)+\beta f_{22}(n))|a,n\rangle \\+(\alpha f_{13}(n)+\beta f_{23}(n))|a,n-1\rangle+(\alpha f_{31}(n)+\beta f_{41}(n))|b,n+1\rangle \\ \nonumber+(\alpha f_{32}(n)+\beta f_{42}(n))|b,n\rangle+(\alpha f_{33}(n)+\beta f_{43}(n))|b,n-1\rangle\},
\end{eqnarray}
where $f_{ij}$'s are given by equation (\ref{53}). Replacing these functions from equation (\ref{53}) the above relation can be simplified into the following general form
\begin{eqnarray}\label{78}
\fl \nonumber |\psi_{\rmtot}(t)\rangle=\sum_{n=0}^{\infty}(c_{a,n}(t)|a,n\rangle+c_{b,n}(t)|b,n\rangle), \qquad \rmwith \\
\nonumber c_{a,n}(t)=\frac{-i}{2}(\alpha-\beta)c_{n-1}\sin(\rmg t\sqrt{n})+c_{n}[(\frac{\alpha+\beta}{2})\cos(\rmg t\sqrt{n})\\ +(\frac{\alpha-\beta}{2})\cos(\rmg t\sqrt{n+1})]-\frac{i}{2}(\alpha+\beta)c_{n+1}\sin(\rmg t\sqrt{n+1})\\
\nonumber c_{b,n}(t)=\frac{-i}{2}(\alpha-\beta)c_{n-1}\sin(\rmg t\sqrt{n})+c_{n}[(\frac{\alpha+\beta}{2})\cos(\rmg t\sqrt{n})\\ \nonumber -(\frac{\alpha-\beta}{2})\cos(\rmg t\sqrt{n+1})]+\frac{i}{2}(\alpha+\beta)c_{n+1}\sin(\rmg t\sqrt{n+1}).
\end{eqnarray}

We can do the trace operation over the basis states of the environment, and obtain the reduced density matrix of the system $\cal S$ as follows
\begin{eqnarray}\label{79}
\nonumber \hat{\rho}^{\cal S}(t)=\sum_{n=0}^{\infty} \langle n|\hat{\rho}^{\rmtot}(t)|n\rangle=\sum_{n=0}^{\infty} \langle n|\psi_{\rmtot}(t)\rangle\langle\psi_{\rmtot}(t)|n\rangle\\=\sum_{n=0}^{\infty}\ (\ |c_{a,n}(t)|^{2}\ |a\rangle\langle a|+|c_{b,n}(t)|^{2}\ |b\rangle\langle b|+c_{a,n}(t)c_{b,n}^{\ast}(t)\ |a\rangle\langle b|+c.c.\ ).
\end{eqnarray}
So, in the basis of the eigenstates of $\sigma_{z}$ (i.e.\ in the basis of the $|a\rangle$ and $|b\rangle$ states) the elements of the reduced density matrix of the system must be given by
\begin{eqnarray}\label{80}
\nonumber \rho^{\cal S}_{12}(t)=\sum_{n=0}^{\infty}c_{a,n}(t)\ c_{b,n}^{\ast}(t)=c_{a,0}\ c_{b,0}^{\ast}+c_{a,1}\ c_{b,1}^{\ast}+c_{a,2}\ c_{b,2}^{\ast}+... \qquad \rmand \\ \rho^{\cal S}_{11}(t)=1-\rho^{\cal S}_{22}(t)=\sum_{n=0}^{\infty}|c_{a,n}(t)|^{2}.
\end{eqnarray}
Replacing $c_{a,n}(t)$ and $c_{b,n}(t)$ from equation (\ref{78}) in the above equation, after some algebra one finds
\begin{eqnarray}\label{81}
\nonumber \rho^{\cal S}_{12}(t)=\gamma f_{0}(t)+\delta f_{1}(t)+\lambda f_{2}(t)+\lambda^{\ast} f_{3}(t)+(\lambda-\lambda^{\ast})f_{4}(t) \qquad \rmand \\
\rho^{\cal S}_{11}(t)=\gamma g_{0}(t)+\delta g_{1}(t)+\lambda g_{2}(t)+\lambda^{\ast} g_{2}^{\ast}(t);
\end{eqnarray}
where in the above equations the coefficients $\gamma,\delta$ and $\lambda$ are given by
\begin{eqnarray}\label{82}
\nonumber \gamma=\frac{1}{4}|\alpha-\beta|^{2} \qquad \rmand \qquad \delta=\frac{1}{4}|\alpha+\beta|^{2}  \qquad \rmand \\
\lambda=\frac{1}{4}(|\alpha|^{2}-|\beta|^{2}+\alpha\beta^{\ast}-\beta\alpha^{\ast}).
\end{eqnarray}
Also the $f_{i}(t)$ and $g_{i}(t)$ functions are given by
\begin{eqnarray}\label{83}
\fl \nonumber f_{0}(t)=\sum_{n=0}^{\infty}(\ |c_{n-1}|^{2}\sin^{2}(\rmg t\sqrt{n})+i[c_{n-1}c_{n}^{\ast}+c_{n-1}^{\ast}c_{n}]\\ \nonumber \times\sin(\rmg t\sqrt{n})\cos(\rmg t\sqrt{n+1})-|c_{n}|^{2}\cos^{2}(\rmg t\sqrt{n+1})\ ), \\
\fl \nonumber f_{1}(t)=\sum_{n=0}^{\infty}(\ |c_{n}|^{2}\cos^{2}(\rmg t\sqrt{n})-i[c_{n}c_{n+1}^{\ast}+c_{n}^{\ast}c_{n+1}]\\ \nonumber \times\cos(\rmg t\sqrt{n})\sin(\rmg t\sqrt{n+1})-|c_{n+1}|^{2}\sin^{2}(\rmg t\sqrt{n+1})\ ), \\
\fl \nonumber f_{2}(t)=\sum_{n=0}^{\infty}(\ -i c_{n-1}c_{n}^{\ast}\sin(\rmg t\sqrt{n})\cos(\rmg t\sqrt{n})-c_{n-1}c_{n+1}^{\ast}\\ \nonumber \times\sin(\rmg t\sqrt{n})\sin(\rmg t\sqrt{n+1})-i c_{n}c_{n+1}^{\ast}\sin(\rmg t\sqrt{n+1})\cos(\rmg t\sqrt{n+1})), \\
\fl \nonumber f_{3}(t)=\sum_{n=0}^{\infty}(\ i c_{n}c_{n-1}^{\ast}\sin(\rmg t\sqrt{n})\cos(\rmg t\sqrt{n})+c_{n+1}c_{n-1}^{\ast}\\ \nonumber \times\sin(\rmg t\sqrt{n})\sin(\rmg t\sqrt{n+1})+i c_{n+1}c_{n}^{\ast}\sin(\rmg t\sqrt{n+1})\cos(\rmg t\sqrt{n+1})), \\
\fl f_{4}(t)=\sum_{n=0}^{\infty}(\ |c_{n}|^{2}\cos(\rmg t\sqrt{n})\cos(\rmg t\sqrt{n+1})), \\
\fl \nonumber g_{0}(t)=\sum_{n=0}^{\infty}(\ |c_{n-1}|^{2}\sin^{2}(\rmg t\sqrt{n})+i[c_{n}c_{n-1}^{\ast}\\ \nonumber-c_{n}^{\ast}c_{n-1}]\sin(\rmg t\sqrt{n})\cos(\rmg t\sqrt{n+1})+|c_{n}|^{2}\cos^{2}(\rmg t\sqrt{n+1})\ ) , \\
\fl \nonumber g_{1}(t)=\sum_{n=0}^{\infty}(\ |c_{n}|^{2}\cos^{2}(\rmg t\sqrt{n})+i[c_{n}c_{n+1}^{\ast}-c_{n}^{\ast}c_{n+1}] \times\cos(\rmg t\sqrt{n})\sin(\rmg t\sqrt{n+1})\\ \nonumber +\ |c_{n+1}|^{2}\sin^{2}(\rmg t\sqrt{n+1}))\quad \rmand\\
\fl \nonumber g_{2}(t)=\sum_{n=0}^{\infty}(\ -i c_{n-1}c_{n}^{\ast}\sin(\rmg t\sqrt{n})\cos(\rmg t\sqrt{n})+c_{n-1}c_{n+1}^{\ast}\\ \nonumber \times\sin(\rmg t\sqrt{n})\sin(\rmg t\sqrt{n+1})+|c_{n}|^{2}\cos(\rmg t\sqrt{n})\cos(\rmg t\sqrt{n+1})\\ \nonumber+i c_{n}c_{n+1}^{\ast}\sin(\rmg t\sqrt{n+1})\cos(\rmg t\sqrt{n+1}))
\end{eqnarray}

\textcolor[rgb]{0.00,0.00,0.00}{In Figure 1 we have used equation (\ref{81}) to plot the evolution of the population inversion $W(t)=\rho_{11}(t)-\rho_{22}(t)$ for the case that the two-level system initially is prepared in the upper level. Here the evolution of the population inversion is characterized by an oscillating envelope, which dominates the fine oscillations which show themselves as collapses and revivals of the atomic inversion in the simpler example of the JCM.}

Now, let us obtain the coherences of the reduced density matrix of the system in another way by using the pointer states of the system and the environment which we obtained in section 3. As we will see, in this way not only we can obtain a closed form for the coherences of the reduced density matrix of the system, but also we can acquire a better understanding regarding the characteristics of decoherence of the central system in our model.

For $|\psi_{\rmtot}(t)\rangle$ given by equation (\ref{73}) the reduced density matrix of the system $\hat{\rho}_{\cal S}(t)$ can be calculated by tracing over the environmental degrees of freedom to obtain
\begin{eqnarray}\label{84}
\fl \nonumber \hat{\rho}_{\cal S}(t)=|\alpha'|^{2}\times|+(t)\rangle\langle +(t)|+|\beta'|^{2}\times|-(t)\rangle\langle -(t)|+\alpha'\beta'^{\ast}\\\fl \ \qquad \times|+(t)\rangle\langle -(t)|\times\langle\Phi_{-}(t)|\Phi_{+}(t)\rangle+\beta'\alpha'^{\ast}\times|-(t)\rangle\langle +(t)|\times\langle\Phi_{+}(t)|\Phi_{-}(t)\rangle.
\end{eqnarray}
So, in an arbitrary basis $|a\rangle$ and $|b\rangle$ of the state of the two-level system generally we have
\begin{eqnarray}\label{85}
\fl \nonumber \rho^{\cal S}_{11}(t)=1-\nonumber \rho^{\cal S}_{22}(t)=|\alpha'|^{2}\times\langle a|+(t)\rangle\langle +(t)|a\rangle+|\beta'|^{2}\times\langle a|-(t)\rangle\langle -(t)|a\rangle+\alpha'\beta'^{\ast}\\ \fl \quad \times\langle a|+(t)\rangle\langle -(t)|a\rangle\times\langle\Phi_{-}(t)|\Phi_{+}(t)\rangle+\beta'\alpha'^{\ast}\times\langle a|-(t)\rangle\langle +(t)|a\rangle\times\langle\Phi_{+}(t)|\Phi_{-}(t)\rangle \\ \fl \nonumber \rmand \qquad
\rho^{\cal S}_{12}(t)=|\alpha'|^{2}\times\langle a|+(t)\rangle\langle +(t)|b\rangle+|\beta'|^{2}\times\langle a|-(t)\rangle\langle -(t)|b\rangle+\alpha'\beta'^{\ast}\\ \fl \quad \nonumber \times\langle a|+(t)\rangle\langle -(t)|b\rangle\times\langle\Phi_{-}(t)|\Phi_{+}(t)\rangle+\beta'\alpha'^{\ast}\times\langle a|-(t)\rangle\langle +(t)|b\rangle\times\langle\Phi_{+}(t)|\Phi_{-}(t)\rangle.
\end{eqnarray}

For the system initially prepared in one of the pointer states $|\pm(t_{0})\rangle$ (i.e.\ for $\alpha'=0$ or $\beta'=0$) the above expressions can be simplified. For example, for the system initially prepared in the $|+(t_{0})\rangle$ state generally we have
\begin{equation}\label{86}
\rho^{\cal S}_{11}(t)=|\langle a|+(t)\rangle|^{2} \qquad \rmand \qquad \rho^{\cal S}_{12}(t)=\langle a|+(t)\rangle.\langle b|+(t)\rangle^{\ast}\ ;
\end{equation}
while for the system initially prepared in the $|-(t_{0})\rangle$ state we have
\begin{equation}\label{87}
\rho^{\cal S}_{11}(t)=|\langle a|-(t)\rangle|^{2} \qquad \rmand \qquad \rho^{\cal S}_{12}(t)=\langle a|-(t)\rangle.\langle b|-(t)\rangle^{\ast}.
\end{equation}

\begin{figure}
  \centering \includegraphics[width=350pt]{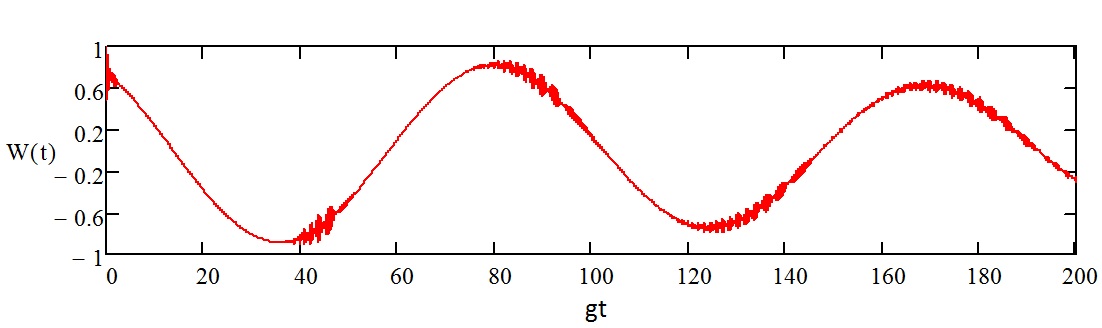}\\
  \caption{Time evolution of the population inversion W(t) for the case that the system initially is prepared in the upper level and for an initial coherent state with $\varphi=\pi/6$ and $\bar{n}=50$.} \label{F1}
\end{figure}

Using equation (\ref{66}) for the pointer states of our spin-boson model, the above equation reads
\begin{eqnarray}\label{88}
\fl \nonumber \rho^{\cal S}_{11}(t)=\cos^{2}(\frac{\varphi}{2}+\frac{\rmg t}{4\sqrt{\bar{n}}}) \ \ \rmand \quad \rho^{\cal S}_{12}(t)=-\frac{i}{2}\sin(\varphi+\frac{\rmg t}{2\sqrt{\bar{n}}}) \ \rmfor \ \ |\psi_{\cal S}(t_{0})\rangle=|+(t_{0})\rangle \\
\fl \rho^{\cal S}_{11}(t)=\sin^{2}(\frac{\varphi}{2}-\frac{\rmg t}{4\sqrt{\bar{n}}}) \ \ \rmand \quad \rho^{\cal S}_{12}(t)=\frac{i}{2}\sin(\varphi-\frac{\rmg t}{2\sqrt{\bar{n}}}) \ \rmfor \ \ |\psi_{\cal S}(t_{0})\rangle=|-(t_{0})\rangle.
\end{eqnarray}
The above expressions for $\rho^{\cal S}_{12}(t)$ basically mean that for the system initially prepared in one of its pointer states, the offdiagonal element of the reduced density matrix of the system should be a sinusoidal function with the frequency of $\frac{\rmg}{2\sqrt{\bar{n}}}$. Also, it must have successive zeros which are apart from each other by $\Delta t=4\pi\sqrt{\bar{n}}/\rmg$.

An examination of $\rho^{\cal S}_{12}(t)$ by plotting its more exact expression, given by equation (\ref{81}), shows good agreement with the above result \emph{only} as long as we have a very large average number of photons in the field. However, for a smaller average number of photons
although we observe the oscillating behavior with the same frequency of $\frac{\rmg}{2\sqrt{\bar{n}}}$, we can clearly observe a decaying envelope which would destroy the offdiagonal element of the reduced density matrix at large times (causing decoherence of the state of the system \emph{even} when the system is initially prepared in one of its pointer states). Also, we observe that this decay is specifically more significant for a smaller average number of photons. Hence, we can guess that the difference between the prediction of equation (\ref{88}) and what we expect from the more exact expression of equation (\ref{81}), for the case that we have a smaller average number of photons, must be due to the fact that in calculating the pointer states of the system we assumed having a large average number of photons in the environment (so that we have a sharp distribution for the coherent state of the field and can assume $\sqrt{n+1}-\sqrt{n}\approx\frac{1}{2\sqrt{\bar{n}}}$). In other words, we guess that \emph{the decoherence of the state of the central system when we start from one of the pointer states of the system must be due to having a limited number of photons in the field.}

In what follows our first goal is to make the appropriate corrections in equation (\ref{88}) so that we can theoretically justify the decoherence of the state of the system when starting from one of the pointer states. Followed by that, we make corrections to the other elements of equation (\ref{85}); and finally, we will use equation (\ref{85}) together with the corrections which we make for having a limited average number of photons, in order to obtain a closed form for $\rho^{\cal S}_{12}(t)$. As we will see, after these corrections our closed form for the offdiagonal element of the reduced density matrix of the system will be in good agreement with the more exact but cumbersome expression of equation (\ref{81}) which we obtained in this section for $\rho^{\cal S}_{12}(t)$.
\textcolor[rgb]{0.00,0.00,0.00}{\subsection{First order corrections due to having a finite average number of photons in the environment}}
In section 3, while obtaining our pointer states of the system and the environment, we assumed having a large average number of photons in the environment; so that we could substitute $\sqrt{n+1}-\sqrt{n}$ by $\frac{1}{2\sqrt{\bar{n}}}$ in our expressions. Now we consider the next order in the Taylor expansion of $\sqrt{n+1}-\sqrt{n}$ about $\bar{n}$, i.e.\ in
\begin{equation}\label{89}
\sum_{n=0}^{\infty}|c_{n}|^{2}\ (\sqrt{n+1}-\sqrt{n})\approx\sum_{n=0}^{\infty}|c_{n}|^{2}\ (\frac{1}{2\sqrt{\bar{n}}}-\frac{(n-\bar{n})}{4\bar{n}^{3/2}}+... ),
\end{equation}
and make the appropriate corrections (due to having a finite average number of photons) in equation (\ref{88}). In fact, by looking at equations (\ref{61}) and (\ref{89}) we notice that \emph{only} at the limit of a large average number of photons, where $\sum_{n=0}^{\infty}|c_{n}|^{2}\ (\sqrt{n+1}-\sqrt{n})\approx\frac{1}{2\sqrt{\bar{n}}}$ is a good approximation and there is no need to consider the next terms in our expansion for $\sqrt{n+1}-\sqrt{n}$\ , the function $G(t)$ will be independent of the states of the environment and pointer states can be realized for the system and the environment, which do not entangle with each other.

We make corrections on $\rho^{\cal S}_{12}(t)$ of equation (\ref{88}) by using the following substitution in our expressions
\begin{equation}\label{90}
e^{\mp it'/2\sqrt{\bar{n}}}\rightarrow\sum_{n=0}^{\infty}|c_{n}|^{2}\ e^{\mp it'(\sqrt{n+1}-\sqrt{n})} \qquad \rmwhere \qquad t'=\rmg t.
\end{equation}
Using equations (\ref{48}) and (\ref{89})
we have
\begin{eqnarray}\label{91}
\nonumber \sum_{n=0}^{\infty}|c_{n}|^{2}\ e^{-it'(\sqrt{n+1}-\sqrt{n})}\approx\sum_{n=0}^{\infty} \frac{e^{-\bar{n}}\ \bar{n}^{n}}{n!}\ e^{-it'(\frac{1}{2\sqrt{\bar{n}}}-\frac{(n-\bar{n})}{4\bar{n}^{3/2}})}\\ =e^{-\bar{n}}\ e^{-\frac{3it'}{4\sqrt{\bar{n}}}}\times\sum_{n=0}^{\infty}
\frac{(\bar{n}\ e^{\frac{it'}{4\bar{n}^{3/2}}})^{n}}{n!}=e^{-\bar{n}}e^{-\frac{3it'}{4\sqrt{\bar{n}}}}\times
\exp(\bar{n}\ e^{\frac{it'}{4\bar{n}^{3/2}}})\\ \nonumber =\exp(\bar{n}\ [e^{\frac{it'}{4\bar{n}^{3/2}}}-1])\ e^{-\frac{3it'}{4\sqrt{\bar{n}}}}=
\exp(\bar{n}\ e^{\frac{it'}{8\bar{n}^{3/2}}}[e^{\frac{it'}{8\bar{n}^{3/2}}}-e^{\frac{-it'}{8\bar{n}^{3/2}}}])\ e^{-\frac{3it'}{4\sqrt{\bar{n}}}};
\end{eqnarray}
which simplifies as
\begin{equation}\label{92}
\sum_{n=0}^{\infty}|c_{n}|^{2}\ e^{-it'(\sqrt{n+1}-\sqrt{n})}\approx\exp(2i\ \bar{n}\ e^{\frac{it'}{8\bar{n}^{3/2}}}\sin(\frac{t'}{8\bar{n}^{3/2}}))\times e^{-\frac{3it'}{4\sqrt{\bar{n}}}}.
\end{equation}
For an average number of photons large enough and times short enough for which $\frac{t'}{\bar{n}^{3/2}}\ll 1$ we can approximate $\sin(\frac{t'}{8\bar{n}^{3/2}})$ by $\frac{t'}{8\bar{n}^{3/2}}$ and $e^{\frac{it'}{8\bar{n}^{3/2}}}$ by $1+\frac{it'}{8\bar{n}^{3/2}}$ in the above equation. In other words, provided $t$ goes to infinity slowly enough to have $\frac{t'}{\bar{n}^{3/2}}\ll 1$ we can write
\begin{eqnarray}\label{93}
\nonumber \sum_{n=0}^{\infty}|c_{n}|^{2}\ e^{-it'(\sqrt{n+1}-\sqrt{n})}\approx\exp(2i\ \bar{n}\ (1+\frac{it'}{8\bar{n}^{3/2}})\times(\frac{t'}{8\bar{n}^{3/2}}))\times e^{-\frac{3it'}{4\sqrt{\bar{n}}}} \qquad \rmor \\
\sum_{n=0}^{\infty}|c_{n}|^{2}\ e^{-it'(\sqrt{n+1}-\sqrt{n})}\approx e^{-\frac{it'}{2\sqrt{\bar{n}}}}\ e^{-t'^{2}/32\bar{n}^{2}}.
\end{eqnarray}
So, to make the appropriate corrections in our expressions we should use the following substitution
\begin{equation}\label{94}
e^{-it'/2\sqrt{\bar{n}}}\rightarrow e^{-it'/2\sqrt{\bar{n}}}\ e^{-t'^{2}/32\bar{n}^{2}}.
\end{equation}
Making the above substitution in the expressions of equation (\ref{88}) for $\rho^{\cal S}_{12}(t)$ we find
\begin{eqnarray}\label{95}
\fl \nonumber \rho^{\cal S}_{12}(t)=-\frac{i}{2}\sin(\varphi+\frac{t'}{2\sqrt{\bar{n}}})\rightarrow -\frac{i}{2}\sin(\varphi+\frac{t'}{2\sqrt{\bar{n}}})\ e^{-t'^{2}/32\bar{n}^{2}} \ \ \rmfor \ \ |\psi_{\cal S}(t_{0})\rangle=|+(t_{0})\rangle \quad \rmand\\
\fl \rho^{\cal S}_{12}(t)=\frac{i}{2}\sin(\varphi-\frac{t'}{2\sqrt{\bar{n}}})\rightarrow \frac{i}{2}\sin(\varphi-\frac{t'}{2\sqrt{\bar{n}}})\ e^{-t'^{2}/32\bar{n}^{2}} \ \ \rmfor \ \ |\psi_{\cal S}(t_{0})\rangle=|-(t_{0})\rangle.
\end{eqnarray}

One interesting aspect of the evolution of coherences given by the above equation is that for $\varphi=0$ and $\varphi=\pi$ no matter whether the system initially is prepared in the $|+(t_{0})\rangle$ state or the $|-(t_{0})\rangle$ state, the evolution of $\rho^{\cal S}_{12}(t)$ is given by $\rho^{\cal S}_{12}(t)=\mp\frac{i}{2}\sin(\frac{t'}{2\sqrt{\bar{n}}})\ e^{-t'^{2}/32\bar{n}^{2}}$ (with the minus sign for $\varphi=0$ and the plus sign for $\varphi=\pi$). Also, if $\varphi=\frac{\pi}{2}$ or $\varphi=\frac{3\pi}{2}$, for both initial pointer states the evolution of $|\rho^{\cal S}_{12}(t)|$ is given by $|\rho^{\cal S}_{12}(t)|=\frac{1}{2}\cos(\frac{t'}{2\sqrt{\bar{n}}})\ e^{-t'^{2}/32\bar{n}^{2}}$. In general, for $\varphi=n\pi/2$ the evolution of $|\rho^{\cal S}_{12}(t)|$ will be the same for both initial pointer states. However, as we will see in the following paragraphs, this does not mean that for $\varphi=n\pi/2$ the evolution of $|\rho^{\cal S}_{12}(t)|$ becomes independent of the initial state of the system.

In Figure 2 we have used equation (\ref{95}) to plot the evolution of $|\rho^{\cal S}_{12}(t)|$ for the case that the system initially is prepared in the $|+(t_{0})\rangle$ state. We also used the more exact expression, given by equation (\ref{81}), to plot the same function. As we see, the correction (due to having a finite average number of photons in the environment) that we made on our initial expression for $\rho^{\cal S}_{12}(t)$, nicely describes the decaying envelope in the evolution of coherences of the reduced system $\cal S$, which is given by the factor $e^{-t'^{2}/32\bar{n}^{2}}$.

\textcolor[rgb]{0.00,0.00,0.00}{$\rho^{\cal S}_{12}(t)$ has also been calculated for the simpler JC-model of quantum optics by Gea-Banacloche \cite{Gea-Banacloche2}. Gea-Banacloche showed that for the case that the two-level system is initially prepared in one of its pointer states the evolution of $|\rho^{\cal S}_{12}(t)|$ is simply given by $|\rho^{\cal S}_{12}(t)|\approx \frac{1}{2}\ e^{-t'^{2}/32\bar{n}^{2}}$ (which is valid for $t'\ll \bar{n}^{3/2}$); as opposed to what we have for our model, given by equation (\ref{95})}.

\textcolor[rgb]{0.00,0.00,0.00}{As we observe, when the system is initially prepared in one of its pointer states the evolution of coherences of our model is characterized by an oscillating envelope, given by $\frac{1}{2}\ |\sin(\varphi+\frac{t'}{2\sqrt{\bar{n}}})|$; while for the simpler JC-model of quantum optics we never observe such ``decayo-sinusoidal" behavior in the evolution of coherences.}
\begin{figure}[t]
  \centering \includegraphics[width=250pt]{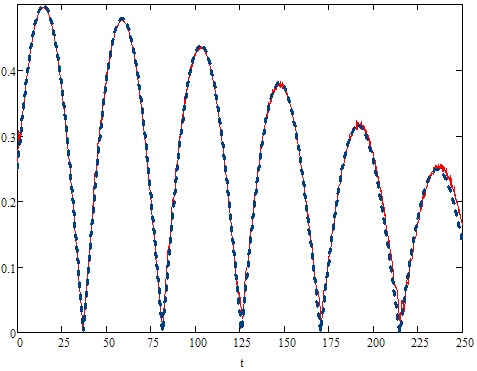}\\
  \caption{Evolution of $|\rho^{\cal S}_{12}(t')|$ (where $t'=\rmg t$) for the case that the system initially is prepared in the $|+(t_{0})\rangle$ state. Here we chose $\varphi=\pi/6$ and $\bar{n}=50$. The curve represented by dashed lines is plotted by using the approximate expression which we obtained from our pointer states, given by equation (\ref{95}). The other curve with solid lines is obtained from the more exact expression of equation (\ref{81}).} \label{F1}
\end{figure}

Coming back to equation (\ref{85}) for the general evolution of coherences of the reduced system, we also need to calculate the expressions for $\langle a|+(t)\rangle\langle -(t)|b\rangle$ and $\langle a|-(t)\rangle\langle +(t)|b\rangle$, as well as the overlap between the pointer states of the environment $\langle\Phi_{-}(t)|\Phi_{+}(t)\rangle$, for our generalized spin boson model.
Using equation (\ref{66}), we can evaluate $\langle a|+(t)\rangle\langle -(t)|b\rangle$ and $\langle a|-(t)\rangle\langle +(t)|b\rangle$ as follows
\begin{eqnarray}\label{96}
\nonumber \langle a|+(t)\rangle\langle -(t)|b\rangle=-i\cos(\frac{\varphi}{2}+\frac{t'}{4\sqrt{\bar{n}}})\cos(\frac{\varphi}{2}-\frac{t'}{4\sqrt{\bar{n}}})  \qquad \rmand \\
\langle a|-(t)\rangle\langle +(t)|b\rangle= i\sin(\frac{\varphi}{2}+\frac{t'}{4\sqrt{\bar{n}}})\sin(\frac{\varphi}{2}-\frac{t'}{4\sqrt{\bar{n}}}).
\end{eqnarray}
However, here also we should make the appropriate correction due to having a finite average number of photons in the environment. Such correction can be made again by using equation (\ref{94}) to obtain
\begin{eqnarray}\label{97}
\nonumber \langle a|+(t)\rangle\langle -(t)|b\rangle=-i\cos(\frac{\varphi}{2}+\frac{t'}{4\sqrt{\bar{n}}})\cos(\frac{\varphi}{2}-\frac{t'}{4\sqrt{\bar{n}}})\ e^{-t'^{2}/64\bar{n}^{2}}  \qquad \rmand \\
\langle a|-(t)\rangle\langle +(t)|b\rangle= i\sin(\frac{\varphi}{2}+\frac{t'}{4\sqrt{\bar{n}}})\sin(\frac{\varphi}{2}-\frac{t'}{4\sqrt{\bar{n}}})\ e^{-t'^{2}/64\bar{n}^{2}}.
\end{eqnarray}

Finally, we use the expressions for $|\Phi_{\pm}(t)\rangle$, given by equations (\ref{70}) and (\ref{71}), in order to calculate the overlap between the pointer states of the environment $\langle\Phi_{-}(t)|\Phi_{+}(t)\rangle$. We have
\begin{equation}\label{98}
\fl \qquad |\Phi_{\pm}(t)\rangle=\sum_{n=0}^{\infty}c_{n}\ e^{\mp\frac{i\rmg t}{2}(\sqrt{n+1}+\sqrt{n})}\ |n\rangle=\sum_{n=0}^{\infty}\frac{e^{-\bar{n}/2}\ \bar{n}^{n/2}\ e^{-in\varphi}}{\sqrt{n!}}\ e^{\mp\frac{i\rmg t}{2}(\sqrt{n+1}+\sqrt{n})}\ |n\rangle.
\end{equation}
As we discussed, for the coherent field with a large average number of photons we can use
\begin{equation}\label{99}
\sqrt{n}\approx\sqrt{\bar{n}}+\frac{(n-\bar{n})}{2\sqrt{\bar{n}}}-\frac{(n-\bar{n})^{2}}{8\bar{n}^{3/2}};
\end{equation}
So, using the above relation and $t'=\rmg t$, equation (\ref{98}) becomes
\begin{eqnarray}\label{100}
\fl \nonumber |\Phi_{\pm}(t)\rangle\approx\sum_{n=0}^{\infty}\frac{e^{-\bar{n}/2}\ \bar{n}^{n/2}\ e^{-in\varphi}}{\sqrt{n!}}\ e^{\mp\frac{it'}{2}(\sqrt{\bar{n}}+\frac{(n+1-\bar{n})}{2\sqrt{\bar{n}}}-\frac{(n+1-\bar{n})^{2}}{8\bar{n}^{3/2}}+\sqrt{\bar{n}}
+\frac{(n-\bar{n})}{2\sqrt{\bar{n}}}-\frac{(n-\bar{n})^{2}}{8\bar{n}^{3/2}})}\ |n\rangle \\
=e^{-\bar{n}/2}\exp(\mp\frac{it'}{2}[\frac{3}{4}\sqrt{\bar{n}}+\frac{3}{4\sqrt{\bar{n}}}-\frac{1}{8\bar{n}^{3/2}}])\times\sum_{n=0}^{\infty}\frac{\bar{n}^{n/2}\ e^{-in\varphi}}{\sqrt{n!}}\\ \nonumber \times \exp(\mp\frac{it'}{2}\{(\frac{n}{\sqrt{\bar{n}}})\times[\frac{3}{2}-\frac{n}{4\bar{n}}]-\frac{n}{4\bar{n}^{3/2}}\})\ |n\rangle.
\end{eqnarray}

For the coherent field and within the approximation that we are using, as we discussed, $\sum_{n} |c_{n}|^{2}\ (n/\bar{n})\approx\sum_{n} |c_{n}|^{2}=1$. Therefore, in the limit of $\bar{n}\rightarrow\infty$ we can replace the expression $[\frac{3}{2}-\frac{n}{4\bar{n}}]$ of the above equation by $\frac{5}{4}$. So, using equation (\ref{48}) we can simplify equation (\ref{100}) to
\begin{equation}\label{101}
|\Phi_{\pm}(t)\rangle\approx\exp(\mp\frac{it'\sqrt{\bar{n}}}{2}[\frac{3}{4}+\frac{3}{4\bar{n}}-\frac{1}{8\bar{n}^{2}}])\times|\ \nu\ \exp(\mp\frac{it'}{2\sqrt{\bar{n}}}[\frac{5}{4}-\frac{1}{4\bar{n}}])\ \rangle.
\end{equation}
So now
\begin{equation}\label{102}
\fl \langle\Phi_{-}(t)|\Phi_{+}(t)\rangle\approx \exp(-it'\sqrt{\bar{n}}\ [\frac{3}{4}+\frac{3}{4\bar{n}}-\frac{1}{8\bar{n}^{2}}])
\times \langle\ \nu\ e^{\frac{it'}{2\sqrt{\bar{n}}}[\frac{5}{4}-\frac{1}{4\bar{n}}]}|\ \nu\ e^{-\frac{it'}{2\sqrt{\bar{n}}}[\frac{5}{4}-\frac{1}{4\bar{n}}]}\ \rangle.
\end{equation}
Using the following formula from quantum optics \cite{Scully} for the scalar product of the coherent states
\begin{equation}\label{103}
    \langle \nu^{\ \prime}|\nu\rangle=\exp[-(|\nu^{\ \prime}|^{2}+
    |\nu|^{2})/2+\nu^{\ \prime\ast}\ \nu].
\end{equation}
equation (\ref{102}) becomes
\begin{equation}\label{104}
\fl \langle\Phi_{-}(t)|\Phi_{+}(t)\rangle\approx\exp(-it'\sqrt{\bar{n}}\ [\frac{3}{4}+\frac{3}{4\bar{n}}-\frac{1}{8\bar{n}^{2}}])\times\exp(\ \bar{n}\
\{e^{\frac{-it'}{\sqrt{\bar{n}}}(\frac{5}{4}-\frac{1}{4\bar{n}})}-1\}).
\end{equation}
Therefore,
\begin{equation}\label{105}
\fl |\langle\Phi_{-}(t)|\Phi_{+}(t)\rangle|^{2}\approx\exp(-4\bar{n}\sin^{2}([\frac{t'}{2\sqrt{\bar{n}}}][\frac{5}{4}-\frac{1}{4\bar{n}}])).
\end{equation}
For an average number of photons large enough and times short enough for which $\frac{t'}{\sqrt{\bar{n}}}\ll 1$ the above expression for the overlap between the pointer states of the environment reduces to
\begin{equation}\label{106}
|\langle\Phi_{-}(t)|\Phi_{+}(t)\rangle|^{2}\approx \exp(-t'^{2}\ [\frac{5}{4}-\frac{1}{4\bar{n}}]^{2})\approx e^{-\frac{25}{16}t'^{2}}.
\end{equation}

Now, using equations (\ref{85}), (\ref{95}), (\ref{97}) and (\ref{104}) we can obtain the following closed form for $\rho^{\cal S}_{12}(t)$, which is valid for $t'\ll \bar{n}^{3/2}$
\begin{eqnarray}\label{107}
\fl \nonumber \rho^{\cal S}_{12}(t)=|\alpha'|^{2}\{\frac{-i}{2}\sin(\varphi+\frac{t'}{2\sqrt{\bar{n}}})\ e^{-t'^{2}/32\bar{n}^{2}}\}+|\beta'|^{2}\{\frac{i}{2}\sin(\varphi-\frac{t'}{2\sqrt{\bar{n}}})\ e^{-t'^{2}/32\bar{n}^{2}}\}\\ \nonumber +\alpha'\beta'^{\ast}\{-i\cos(\frac{\varphi}{2}+\frac{t'}{4\sqrt{\bar{n}}})\cos(\frac{\varphi}{2}-\frac{t'}{4\sqrt{\bar{n}}})\ e^{-t'^{2}/64\bar{n}^{2}}\}\times \\ \exp(-it'\sqrt{\bar{n}}\ [\frac{3}{4}+\frac{3}{4\bar{n}}-\frac{1}{8\bar{n}^{2}}]) \times\exp(\ \bar{n}\
\{e^{\frac{-it'}{\sqrt{\bar{n}}}(\frac{5}{4}-\frac{1}{4\bar{n}})}-1\})
\\ \nonumber +\beta'\alpha'^{\ast}\{i\sin(\frac{\varphi}{2}+\frac{t'}{4\sqrt{\bar{n}}})\sin(\frac{\varphi}{2}-\frac{t'}{4\sqrt{\bar{n}}})\ e^{-t'^{2}/64\bar{n}^{2}}\}\times\\ \nonumber \exp(it'\sqrt{\bar{n}}\ [\frac{3}{4}+\frac{3}{4\bar{n}}-\frac{1}{8\bar{n}^{2}}]) \times\exp(\ \bar{n}\
\{e^{\frac{it'}{\sqrt{\bar{n}}}(\frac{5}{4}-\frac{1}{4\bar{n}})}-1\}).
\end{eqnarray}

In Figure 3 we used equation (\ref{107}) to plot the short time evolution of $|\rho^{\cal S}_{12}(t)|$ for a case that the system initially is \emph{not} prepared in one of its pointer states. We also used the more exact expression, given by equation (\ref{81}), to plot the same function. As we see from this figure, equation (\ref{107}) serves as a good approximation in closed form for the more exact relation, as long as we are not considering long times \footnote{\textcolor[rgb]{0.00,0.00,0.00}{For longer times, the approximate equation (\ref{107}) shows internal oscillations in the evolution of $|\rho^{\cal S}_{12}(t)|$ which are misplaced compared to those of the plot which we obtain from the more exact expression of equation (\ref{81}). However, the envelopes still do coincide with each other with great precession.}}.
\begin{figure}
  \centering \includegraphics[width=350pt]{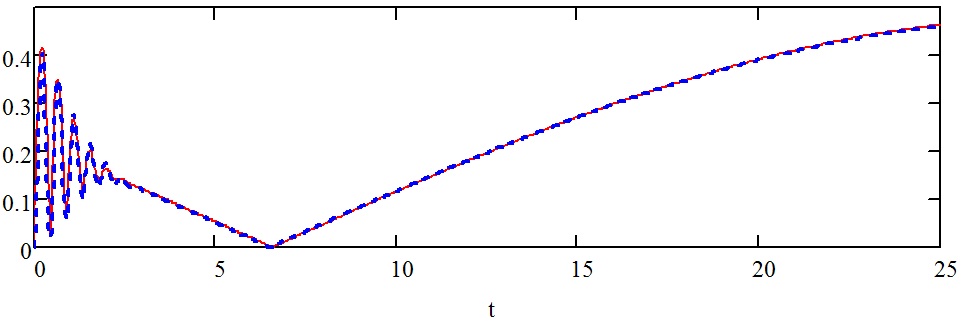}\\
  \caption{Short time evolution of $|\rho^{\cal S}_{12}(t')|$ for the case that the system initially is prepared in the lower state.
  Here we chose $\varphi=\pi/6$ and $\bar{n}=50$. The curve represented by dashed lines is plotted by using the approximate expression which we obtained from our pointer states, given by equation (\ref{107}). The other curve with solid lines is obtained from the more exact expression of equation (\ref{81}).} \label{F1}
\end{figure}

As we can see from equations (\ref{106}) and (\ref{107}), at sufficiently short times $t'\ll \sqrt{\bar{n}}$ the decay of the first two terms is characterized by the decaying factor $e^{-t'^{2}/32\bar{n}^{2}}$, while the decay of the other two terms is characterized by the much faster-decaying term due to the overlap between the pointer states of the environment $\langle\Phi_{-}(t)|\Phi_{+}(t)\rangle$ which is proportional to the factor $e^{-\frac{25}{32}t'^{2}}$. This fact clearly shows why indeed we should generally expect a much slower decoherence of the state of the system when the system is initially prepared in one of its pointer states, compared to the case that the system initially is \emph{not} in any of its pointer states. Also, equation (\ref{107}) shows that for $\varphi=n\pi/2$ and if the system initially is \emph{not} prepared in one of its pointer states, the evolution of coherences of the central system would not be independent of the initial state of the system (just unlike the case that the system is initially prepared in one of its pointer states). However, since the last two terms of equation (\ref{107}) vanish much faster than the first two terms, at larger times and for $\varphi=n\pi/2$ we expect the evolution of coherences of the central system to be independent of the initial state of the system.

In this section we calculated the offdiagonal element of the reduced density matrix of the system in the basis of eigenstates of the $\hat{\sigma}_{z}$ operator. However, we could equally study the decoherence of the state of the central system in the basis of the $|\pm(t_{0})\rangle$ states.

From equation (\ref{66}) it is clear that for $t'\ll \sqrt{\bar{n}}$ the pointer states of the system almost are time-independent; and they can be approximated by $|\pm(t_{0})\rangle$. So, in this limit the evolution of an arbitrary initial state of the central system $(\alpha'\ |+(t_{0})\rangle+\beta'\ |-(t_{0})\rangle)$ in contact with an initial coherent field $|\nu\rangle$ is approximately given by $|\psi^{\rmtot}(t)\rangle=\alpha'\ |+(t_{0})\rangle|\Phi_{+}(t)\rangle+\beta'\ |-(t_{0})\rangle|\Phi_{-}(t)\rangle$. Therefore, we would have
\begin{eqnarray}\label{108}
\fl \nonumber \hat{\rho}_{\cal S}(t)=|\alpha'|^{2}\ |+(t_{0})\rangle\langle +(t_{0})|+|\beta'|^{2}\ |-(t_{0})\rangle\langle -(t_{0})|\\ \fl \quad +\alpha'\beta'^{\ast}\ |+(t_{0})\rangle\langle -(t_{0})|\times\langle\Phi_{-}(t)|\Phi_{+}(t)\rangle+\beta'\alpha'^{\ast}\ |-(t_{0})\rangle\langle +(t_{0})|\times\langle\Phi_{+}(t)|\Phi_{-}(t)\rangle.
\end{eqnarray}
For this short range of times the evolution of the pointer states of the environment can be approximated by equation (\ref{101}). So, in the $|\pm(t_{0})\rangle$ basis and for $t'\ll \sqrt{\bar{n}}$ we must have
\begin{eqnarray}\label{109}
\fl \nonumber \rho^{\cal S}_{12}(t)=\alpha'\beta'^{\ast}\langle\Phi_{-}(t)|\Phi_{+}(t)\rangle \\ \approx\alpha'\beta'^{\ast}\exp(-it'\sqrt{\bar{n}}\ [\frac{3}{4}+\frac{3}{4\bar{n}}-\frac{1}{8\bar{n}^{2}}]) \times\exp(\ \bar{n}\
\{e^{\frac{-it'}{\sqrt{\bar{n}}}(\frac{5}{4}-\frac{1}{4\bar{n}})}-1\}).
\end{eqnarray}
Finally, using equation (\ref{106}) we find that for $t'\ll \sqrt{\bar{n}}$
\begin{equation}\label{110}
|\rho^{\cal S}_{12}(t)|^{2}\approx|\alpha'\beta'|^{2}\exp(-t'^{2}\ [\frac{5}{4}-\frac{1}{4\bar{n}}]^{2})\approx|\alpha'\beta'|^{2}\ e^{-\frac{25}{16}t'^{2}}.
\end{equation}
Hence, in the basis of the $|\pm(t_{0})\rangle$ states the short-time decoherence of the state of the central system is characterized by the fast-decaying factor
$e^{-\frac{25}{32}t'^{2}}$ when the system initially is \emph{not} prepared in one of its pointer states; while in this basis the pointer states of the system almost do not decohere within short times.\\

\section{Summary and conclusions}

Considering a single-mode quantized field in exact resonance with the tunneling matrix element of the system, we obtained the time-evolution operator for our model. Using this time-evolution operator then we calculated the pointer states of the system and the environment for the case that the environment is initially prepared in the coherent state with a large average number of photons. Most importantly, we observed that for our spin-boson model represented by the Hamiltonian of equation (\ref{1}) the pointer states of the system turn out to become time-dependent; as opposed to the pointer states of a simplified spin-boson model (with $\hat{H}_{\cal S}$ proportional to $\hat{\sigma}_{z}$ rather than $\hat{\sigma}_{x}$) for which $[\hat{H}_{\cal S},\hat{H}_{\rm int}]=0$. The simplified model has often been used in the context of quantum information and quantum computation to gain some insights regarding the decoherence of a single qubit \cite{Ekert,Unruh,Reina}. However, in most of the practical situations different noncommutable perturbations may exist in the total Hamiltonian of a realistic system-environment model which would result in having time-dependent pointer states for the system \cite{paper1}. Indeed, the authors believe that the fact that the pointer states of a system generally are time-dependent and may evolve with time has not been seriously acknowledged in the context of quantum computation and quantum information. In specific, in the context of quantum error correction \cite{Laflamme,Nielsen} it is often assumed that the premeasurement by the environment does not change the initial pointer states of the system. In other words, quantum ``nondemolition" premeasurement by the environment is often assumed \cite{Laflamme,Nielsen}; as is also assumed in Von Neumann's scheme of measurement \cite{Neuman,Schlosshauer1}. Also, in the context of Decoherence-Free-Subspaces (DFS) theory the models which often are studied either contain self-Hamiltonian for the system which commutes with the interaction between the system and the environment, or it is assumed that we are in the \emph{quantum measurement limit} \footnote{In the \emph{quantum measurement limit} the interaction between the system and the environment is so strong as to dominate the evolution of the system $\hat{H}\approx \hat{H}_{\rm int}$. Also in the \emph{quantum limit of decoherence} the Hamiltonian for the system almost dominates the interaction between the system and the environment as well as the self-Hamiltonian of the environment $\hat{H}\approx \hat{H}_{\cal S}$.} or in the \emph{quantum limit of decoherence} \cite{Ekert,90,91,93}. However, all of these assumptions are in fact a big simplification of the problem; since, as we discussed in paper $\sf I$, they completely exclude the possibility of having pointer states for the system which may depend on time \cite{paper1}.

Another interesting point in obtaining the pointer states of the system and the environment for our model was the realization of the fact that \emph{only} in the limit of a large average number of photons can we have a set of (time-dependent) pointer states for the system. In other words, unless we have a sufficiently large average number of photons which can make a sharp distribution function
for the state of the electromagnetic field, there is always some degree of entanglement between the states of the system and the environment (see equations (\ref{61}) and (\ref{89})) and the pointer states of measurement cannot be realized at all.

We also showed that at $t=(2n+1)\pi\sqrt{\bar{n}}/\rmg \ (\rmwith\ n=0,1,2,...)$ the $|\pm(t)\rangle$ pointer states of the system coincide with each other and hence, whatever is the initial state of the system, at these specific times the states of the system and the environment are not entangled with each other and the system can be represented by a well-defined state of its own. Using our pointer states, we also obtained a closed form for the offdiagonal element of the reduced density matrix of the system and studied the decoherence of the central system in our model. We showed that for the case that the system initially is prepared in one of its pointer states, the offdiagonal element of the reduced density matrix of the system will be \textcolor[rgb]{0.00,0.00,0.00}{a \emph{sinusoidal function} with a slow decaying envelope which is characterized by a decay time proportional to $\bar{n}$ (through a decoherence factor calculated as $e^{-\rmg^{2} t^{2}/32\bar{n}^{2}}$)}; while for the case that the system initially is not prepared in one of its initial pointer states, it will experience a fast decoherence within a time of order $1/\rmg$. The ``decayo-sinusoidal" evolution of coherences (figure 2) which we observe in our model and for the case that the system initially is prepared in one of its pointer states is a new form of decoherence which cannot be observed in the somewhat similar Jaynes-Cummings model of quantum optics \cite{Gea-Banacloche2}.

It will be interesting to generalize this study to the case that the environment is not merely represented by a single-mode bosonic field; and consider some classes of spectral densities for the environment. Also, for the spin-boson model represented by the Hamiltonian of equation (\ref{1}) at least in principle one should be able to obtain the pointer states of the system and the environment in some nonresonance regimes and for the single-mode quantized field.

To further demonstrate the generality and usefulness of our method of obtaining pointer states, in another article \cite{paper4} we will obtain the time-dependent pointer states of the system and the environment for the quantized atom-field model and in some nonresonance regimes.

\appendix
\section*{Appendix A}
\renewcommand{\theequation}{A-\arabic{equation}}
\setcounter{equation}{0}

In this section we introduce a measure for degree of entanglement between the states of the system and those of the environment, and by calculating this measure for the initial pointer states of our model (which are given by equation (\ref{63})) we show that our result for the pointer states of the system and the environment is valid over a length of time which is proportional to $\bar{n}$, the average number of photons in the field; since as we will see, only within up to this range of times our pointer states of the system and the environment can stay separated and will not considerably entangle with the states of another subsystem.

For the global state of the system and the environment, given by equation (\ref{310}), we have
\begin{eqnarray}\label{A-1}
\fl \nonumber |\psi^{\rm tot}(t)\rangle=|\textbf{A}(t)\rangle\ |a\rangle+|\textbf{B}(t)\rangle\ |b\rangle \\ =|\textbf{A}^{\prime}(t)\rangle\ [G(t)|a\rangle]+|\textbf{B}(t)\rangle\ |b\rangle;\ \ \rmwhere\ \ |\textbf{A}^{\prime}(t)\rangle=\frac{|\textbf{A}(t)\rangle}{G(t)}.
\end{eqnarray}
For our pointer states (given by $|\pm(t)\rangle=\calN_{\pm}\ \{G_{\pm}(t)|a\rangle+|b\rangle\}$) we can define a degree of entanglement through the following relation
\begin{equation}\label{A-2}
q(t)=\frac{\langle\textbf{A}(t)|G_{\pm}(t)\textbf{B}(t)\rangle}{|\textbf{A}(t)|^{2}};
\end{equation}
which also is equal to
\begin{equation}\label{A-3}
q(t)=\frac{\langle\textbf{A}^{\prime}(t)|\textbf{B}(t)\rangle}{|\textbf{A}^{\prime}(t)|^{2}}.
\end{equation}
The above function basically is the overlap between the vectors $|\textbf{A}^{\prime}(t)\rangle$ and $|\textbf{B}(t)\rangle$ of equation (\ref{A-1}); which is \emph{normalized} to the unity, since for our pointer states of the system we have $|\textbf{A}(t)\rangle=G_{\pm}(t)|\textbf{B}(t)\rangle$. From equation (\ref{A-1}) it is clear that for perfect pointer states, where there is no entanglement between the states of the system and the environment, $q(t)$ must always remain equal to the unity; all throughout the evolution of the system and the environment (i.e.\ $|\textbf{A}^{\prime}(t)\rangle$ and $|\textbf{B}(t)\rangle$ must perfectly coincide with each other, in which case the states of the system and the environment in equation (A-1) will not entangle with each other; and we will have pointer states for the system which are given by equation (11)). Only in this case the states of the system and the environment in equation (\ref{A-1}) will always stay separated and one can assign each of the two subsystems with well-defined states of their own.

Our goal is to calculate our measure of entanglement $q(t)$ for the pointer states of the system and the environment which we obtained for our model in this paper; and to study its evolution with time. For the global global state of the system and the environment, given by equation (\ref{A-1}), one can calculate the reduced density matrix of the system as
\begin{equation}\label{A-5}
\fl \hat{\rho}^{\cal S}(t)=|a\rangle\langle a|\langle\textbf{A}(t)|\textbf{A}(t)\rangle+|b\rangle\langle b|\langle\textbf{B}(t)|\textbf{B}(t)\rangle+|a\rangle\langle b|\langle\textbf{B}(t)|\textbf{A}(t)\rangle+|b\rangle\langle a|\langle\textbf{A}(t)|\textbf{B}(t)\rangle.
\end{equation}
Therefore,
\begin{equation}\label{A-6}
\rho^{\cal S}_{12}(t)=\langle\textbf{B}(t)|\textbf{A}(t)\rangle;
\end{equation}
where in the above equation $\rho^{\cal S}_{12}(t)$ is the offdiagonal element of the reduced density matrix of the two-level system in the $|a\rangle$ and $|b\rangle$ basis.

From equation (\ref{A-6}) we can easily see that if the system initially is prepared in one of its initial pointer states; i.e.\ if $|\psi^{\cal S}(t_{0})\rangle=|\pm(t_{0})\rangle$, then we have:
\begin{equation}\label{A-7}
\fl \langle\textbf{A}(t)|G_{\pm}(t)\textbf{B}(t)\rangle=G_{\pm}^{\ast}(t)\langle\textbf{B}(t)|\textbf{A}(t)\rangle=G_{\pm}^{\ast}(t)\rho^{\cal S}_{12}(t)
\end{equation}
Therefore,
\begin{equation}\label{A-8}
q(t)=\frac{\langle\textbf{A}(t)|G_{\pm}(t)\textbf{B}(t)\rangle}{|\textbf{A}(t)|^{2}}=\frac{G_{\pm}^{\ast}(t)\rho^{\cal S}_{12}(t)}{|\textbf{A}(t)|^{2}};\quad \rmor
\end{equation}
\begin{equation}\label{A-8.5}
|q(t)|=\frac{|\rho^{\cal S}_{12}(t)|}{|\textbf{A}(t)|\times|\textbf{B}(t)|}=\frac{|\rho^{\cal S}_{12}(t)|}{\sqrt{\rho^{\cal S}_{11}(t)}\times\sqrt{1-\rho^{\cal S}_{11}(t)}}.
\end{equation}
In calculating the above equations we must keep in mind that $\rho^{\cal S}_{11}(t)$ and $\rho^{\cal S}_{12}(t)$ must be calculated in the basis of the $|a\rangle$ and $|b\rangle$ basis states; and also they must be calculated for the case that the system initially is prepared in one of its initial pointer states, given by equation (\ref{63}).

For our spin-boson model and for example for the case that the system initially is prepared in the $|+(t_{0})\rangle$ state, from equations (\ref{88}), (\ref{94}) and (\ref{95}) we had
\begin{equation}\label{A-10}
\fl \rho^{\cal S}_{11}(t)=\cos^{2}(\frac{\varphi}{2}+\frac{t'}{4\sqrt{\bar{n}}})\ e^{-t'^{2}/32\bar{n}^{2}} \ \ \rmand \ \ |\rho^{\cal S}_{12}(t)|=\frac{1}{2}|\sin(\varphi+\frac{t'}{2\sqrt{\bar{n}}})|\ e^{-t'^{2}/32\bar{n}^{2}}.
\end{equation}
Therefore,
\begin{equation}\label{A-11}
|q(t)|=\frac{|\rho^{\cal S}_{12}(t)|}{\sqrt{\rho^{\cal S}_{11}(t)}\times\sqrt{1-\rho^{\cal S}_{11}(t)}}\simeq1\quad \rm if\ and\ only\ if\quad t'\ll \bar{n}
\end{equation}
One can easily verify that for the case that the system initially is prepared in the $|-(t_{0})\rangle$ state also, we would have the same result for the degree of entanglement between the states of the system and the environment. These results basically indicate that our result for the pointer states of the system and the environment is valid over a length of time which is proportional to $\bar{n}$, the average number of photons in the field; since within times of the order of $\bar{n}/\rmg$ the degree of entanglement, calculated for our pointer states, will stay close to the unity and our calculated pointer states will be immune to entanglement.




\section*{References}
\begin{thebibliography}{99}
\bibitem{Schlosshauer1}
Schlosshauer M 2007 \emph{Decoherence and the Quantum-to-Classical
Transition} (Berlin Heidelberg: Springer);\\
\nonumber Joos E, Zeh H D, Kiefer C, Giulini D, Kupsch J and Stamatescu I O 2003 \emph{Decoherence and the Appearance of the Classical World}
(Springer)
\bibitem{Zurek1}
Zurek W H 1981 \emph{Phys. Rev.} D \textbf{24} 1516;\\
\nonumber Zurek W H 2003 \emph{Rev. Mod. Phys.} \textbf{75} 715
\bibitem{Zurek2}
Zurek W H 1982 \emph{Phys. Rev.} D \textbf{26} 1862
\bibitem{paper1}
Daneshvar H and Drake G W F, e-print arXiv:1104.4405v1 [quant-ph].\\
\noindent A short version of the above article is submitted for publication at \emph{Phys. Rev.} A.
\bibitem{Gea-Banacloche}
\nonumber Gea-Banacloche J 1991 \emph{Phys. Rev.} A \textbf{44} 5913
\bibitem{Gea-Banacloche2}
Gea-Banacloche J 1992 \emph{Phys. Rev.} A \textbf{46} 7307\\
Gea-Banacloche J 1992 \emph{Optics Communications} \textbf{88} 531-550
\bibitem{Knight}
Phoenix S J and Knight P L 1991 \emph{Phys. Rev.} A \textbf{44} 6023
\bibitem{Leggett}
Leggett A J, Chakravarty S, Dorsey A T, Fisher M P A and Garg A 1987 \emph{Rev. Mod. Phys.}
\textbf{59} 1-85
\bibitem{Weiss}
Weiss U 1999 \emph{Quantum Dissipative Systems} (Singapore: World Scientific)
\bibitem{Scully}
Scully M O and Zubairy M S 1997 \emph{Quantum Optics} (Cambridge: Cambridge University Press)
\bibitem{Drake}
Meschede D and Schenzle A, in \emph{Handbook of Atomic, Molecular, and Optical Physics}, Editted by Drake G W F 2005 (Springer);\\
Narozhny N B, Sanchez-Mondragon J J, and Eberly J H 1981 \emph{Phys. Rev.} A \textbf{23} 236
\bibitem{Feynman}
Feynman R P and Vernon F L 1963 \emph{Ann. Phys.} (N.Y.) \textbf{24} 118
\bibitem{Duan}
Duan L M and Guo G C 1998 \emph{Phys. Rev.} A \textbf{57} 737
\bibitem{Eberly}
Allen L and Eberly J H 1987 \emph{Optical Resonance and Two-Level Atoms} (New York: Dover Publications)
\bibitem{Ekert}
Palma G M, Suominen K A and Ekert A K 1996 \emph{Proc. R. Soc. Lond.} A \textbf{452} 567
\bibitem{Unruh}
Unruh W G 1995 \emph{Phys. Rev.} A \textbf{51} 992
\bibitem{Reina}
Reina J H, Quiroga L and Johnson N F 2002 \emph{Phys. Rev.} A \textbf{65} 032326
\bibitem{Laflamme}
Knill E, Laflamme R, Ashikhmin A, Barnum H, Viola L and Zurek W 2002 \emph{Introduction to quantum error correction}, \emph{LA Science} \textbf{27} 188-225
\bibitem{Nielsen}
Nielsen M A and Chuang I L 2000 \emph{Quantum Computation and Qunatum Information} (Cambridge: Cambridge University Press)
\bibitem{Neuman}
Von Neumann J 1932 \emph{Mathematische Grundlagen der Quantenmechanik} (Springer: Berlin)
\bibitem{90}
Lidar D A, Chuang I L and Whaley K B 1998 \emph{Phys. Rev. Lett.} \textbf{81} 2594
\bibitem{91}
Zanardi P and Rasetti M 1997 \emph{Phys. Rev. Lett} \textbf{79} 3306
\bibitem{93}
Zanardi P 1998 \emph{Phys. Rev.} A \textbf{57} 3276
\bibitem{paper4}
Daneshvar H and Drake G W F, e-print arXiv:1104.3949v1 [quant-ph]. Also submitted for publication at \emph{J. Phys.} A.

\end{thebibliography}
\end{document}